\documentclass[aps,twocolumn]{revtex4}

\usepackage{graphics}
\usepackage{epsfig}

\usepackage{amsmath}
\usepackage{amssymb}
\usepackage{stmaryrd}

\def\ov#1{\overline{#1}}

\def\wt#1{\widetilde{#1}}
\def\vb#1{\mbox{\boldmath$#1$}}
\def\pd#1#2{\frac{\partial #1}{\partial #2}}

\def\wh#1{\widehat{#1}}
\def\bdot{\,\vb{\cdot}\,}
\def\btimes{\,\vb{\times}\,}

\def\exd{{\sf d}}

\newcommand{\bc}{\begin{center}}
\newcommand{\ec}{\end{center}}
\newcommand{\bt}{\begin{tabbing}}
\newcommand{\et}{\end{tabbing}} 
\newcommand{\be}{\begin{eqnarray*}}
\newcommand{\ee}{\end{eqnarray*}}

\begin{document}

\title{Particle and guiding-center orbits in crossed electric and magnetic fields}

\author{Alain J.~Brizard$^{a}$}
\affiliation{Department of Physics, Saint Michael's College, Colchester, VT 05439, USA \\ $^{a}$Author to whom correspondence should be addressed: abrizard@smcvt.edu}

\begin{abstract}
The problem of the charged-particle motion in crossed electric and magnetic fields is investigated, and the validity of the guiding-center representation is assessed in comparison with the exact particle dynamics. While the magnetic field is considered to be straight and uniform, the (perpendicular) radial electric field is nonuniform. The Hamiltonian guiding-center theory of charged-particle motion is presented for arbitrary radial electric fields, and explicit examples are provided for the case of a linear radial electric field.
\end{abstract}

\date{\today}

\maketitle

\section{Introduction}

The importance of radial electric fields in rotating magnetized plasmas has been a topic of great interest for a few decades \cite{Hahm:1996,Hahm:2002,Peeters:2009,Burrell:2020} because of the role of sheared $E\times B$ flows in the stabilization of turbulent magnetized plasmas. The guiding-center analysis for particle dynamics in the presence of background (equilibrium) radial electric fields \cite{Brizard:1995,Cary_Brizard:2009}, and its extension to the gyrokinetic analysis of turbulent magnetized plasmas \cite{Dimits:2010,Sugama:2011,Belli_Candy:2018,Frei_etal:2020,Wang_etal:2021}, has also been a topic of great interest. The case of rotating magnetized plasmas due to the presence of radial electric fields continues to attract attention \cite{White:2018,Joseph:2021}.

In the present work, we apply the Hamiltonian guiding-center theory developed in Ref.~\cite{Brizard:1995} for general magnetic geometry to the problem of charge-particle motion in the presence of a uniform magnetic field ${\bf B} = B_{0}\,\wh{\sf z}$ and a nonuniform electric field ${\bf E} = -\,\nabla\Phi$. Using the magnetic coordinates $(\psi,\theta,z)$, the magnetic field ${\bf B} = \nabla\btimes{\bf A}$ can be represented in terms of the magnetic vector potential ${\bf A} = \psi\,\nabla\theta$, where the magnetic flux is expressed in terms of cylindrical coordinates as $\psi(r) = B_{0}r^{2}/2$, while the electric potential $\Phi = \Phi(\psi)$ is assumed to be a flux function. While the standard assumption is that the $E\times B$ rotation frequency $\omega_{E}(\psi) \equiv c\,\Phi^{\prime}(\psi) \ll \Omega_{0}$ is small compared to the gyrofrequency $\Omega_{0} = qB_{0}/mc$ (for a particle of mass $m$ and charge $q$), this does not require the $E\times B$ velocity $r\,\omega_{E}(\psi)$ to be small compared to the thermal velocity \cite{Dimits:2010,Joseph:2021}, especially in the edge of a confined magnetized plasma.

The remainder of the paper is organized as follows. First, the dynamics of a charged particle moving in the two-dimensional plane perpendicular to a constant magnetic field under the influence of a nonlinear radial electric field (Sec.~\ref{sec:particle_nonlinear}) and a linear radial electric field (Sec.~\ref{sec:particle_linear}) is discussed, where explicit results are presented for the linear case. Next, the guiding-center analysis for particle dynamics in a nonlinear radial electric field is presented in Sec.~\ref{sec:nonlinear_gc} while explicit guiding-center results are presented in Sec.~\ref{sec:linear_gc} for a linear radial electric field.

\section{\label{sec:particle_nonlinear}Particle Lagrangian Dynamics in Nonlinear Radial Electric Field}

The two-dimensional motion of a charged particle in the plane perpendicular to the magnetic field is represented by the particle Lagrangian expressed in magnetic coordinates as
\begin{eqnarray}
L &=& \left( \frac{q}{c}\,{\bf A} \;+\; m\,{\bf v}\right)\bdot\dot{\bf x} \;-\; \left( \frac{m}{2}\,|{\bf v}|^{2} \;+\; q\,\Phi\right) \nonumber \\
 &=& \frac{q}{c}\,{\bf A}\bdot\dot{\bf x} \;+\; \frac{m}{2}\,|\dot{\bf x}|^{2} \;-\; q\,\Phi \nonumber \\
 &=&\frac{q}{c}\,\psi\,\dot{\theta} \;+\; \frac{q\psi}{c\Omega_{0}}\left( \frac{\dot{\psi}^{2}}{4\,\psi^{2}} + \dot{\theta}^{2}\right) \;-\; q\,\Phi(\psi),
\label{eq:Lag_particle}
\end{eqnarray}
where a dot denotes a time derivative and the particle's velocity
\begin{equation}
{\bf v} \;\equiv\;  \dot{\bf x} \;=\; \dot{\psi}\;\pd{\bf x}{\psi} \;+\; \dot{\theta}\;\pd{\bf x}{\theta} \;=\;  \sqrt{\frac{2\psi}{B_{0}}} \left( \frac{\dot{\psi}}{2\psi}\,\wh{r} \;+\; \dot{\theta}\,\wh{\theta}\right) 
\label{eq:x_dot}
\end{equation}
is expressed in the plane perpendicular to the magnetic field, with Jacobian $\wh{\sf z}\bdot(\partial{\bf x}/\partial\psi\btimes\partial{\bf x}/\partial\theta) = 1/B_{0}$. Since the Lagrangian \eqref{eq:Lag_particle} is independent of the azimuthal angle $\theta$, the azimuthal canonical angular momentum
\begin{equation}
p_{\theta} \;\equiv\; \pd{L}{\dot{\theta}} \;=\;  \frac{q}{c}\,\psi \left( 1 \;+\; 2\,\dot{\theta}/\Omega_{0}\right)
\label{eq:p_theta}
\end{equation}
is a constant of motion (i.e., $dp_{\theta}/dt = \partial L/\partial\theta = 0$). By using the initial conditions $(\psi_{0},\dot{\theta}_{0} = \omega_{0})$, we may write the azimuthal canonical angular momentum as
\begin{equation}
p_{\theta} \;=\;  \frac{q}{c}\,\psi_{0} \left( 1 \;+\; 2\,\omega_{0}/\Omega_{0}\right) \;\equiv\; \frac{q}{c}\,\ov{\psi}_{0},
\label{eq:p_theta0}
\end{equation}
and, thus, the azimuthal angular velocity is
\begin{equation}
\dot{\theta}(\psi) \;=\; \frac{\Omega_{0}}{2} \left( \frac{\ov{\psi}_{0}}{\psi} - 1 \right),
\label{eq:theta_dot}
\end{equation}
which indicates that reversal is possible if $\psi$ crosses $\ov{\psi}_{0}$.

Next, we construct the Routh-Lagrange function (or Routhian \cite{Brizard:2015}) in order to obtain a reduced Lagrangian formulation of the $\psi$-dynamics:
\[ {\sf R}(\psi,\dot{\psi}) \;\equiv\; L - \dot{\theta}\,\partial L/\partial\dot{\theta}, \]
and we obtain
\begin{equation}
{\sf R}(\psi,\dot{\psi}) \;=\; \frac{q}{c\Omega_{0}} \left(\frac{\dot{\psi}^{2}}{4\,\psi}\right) \;-\; V(\psi),
\label{eq:Routh}
\end{equation} 
where the effective potential is
\begin{equation}
V(\psi) \;=\; q\,\Phi(\psi) \;+\;  \frac{q \Omega_{0}}{4c\;\psi}\left(\ov{\psi}_{0} - \psi\right)^{2}.
\label{eq:V_psi}
\end{equation}
Hence, the Routh-Euler-Lagrange equation 
\[ \frac{d}{dt}\left(\pd{{\sf R}}{\dot{\psi}}\right) \;=\; \pd{\sf R}{\psi} \]
yields the nonlinear second-order ordinary differential equation for the normalized magnetic flux $\chi \equiv \psi/\ov{\psi}_{0}$:
\begin{equation}
\chi^{\prime\prime} \;=\; \frac{1}{2\chi} \left[ (\chi^{\prime})^{2} \;+\frac{}{} \left(1 \;-\; \chi^{2}\right)\right] \;-\; 2\epsilon\,\chi\,\nu(\chi),
\label{eq:chi_pp}
\end{equation}
where a prime denotes a derivative with respect to the normalized time $t^{\prime} \equiv \Omega_{0}\,t$ and the dimensionless parameter 
\begin{equation}
\epsilon \;\equiv\; c\,\Phi^{\prime}(\ov{\psi}_{0})/\Omega_{0} 
\label{eq:epsilon}
\end{equation}
denotes the ratio of the $E\times B$ azimuthal frequency $\omega_{E}(\ov{\psi}_{0}) = c\Phi^{\prime}(\ov{\psi}_{0})$ at $\psi = \ov{\psi}_{0}$ to the gyrofrequency $\Omega_{0}$, and we define the function
\begin{equation}
\nu(\chi) \;\equiv\; \Phi^{\prime}(\ov{\psi}_{0}\chi)/\Phi^{\prime}(\ov{\psi}_{0}),
\label{eq:nu}
\end{equation}
which equals one if $\Phi(\psi)$ is a linear function of $\psi$.

We note that Eq.~\eqref{eq:chi_pp} possesses an energy conservation law, with normalization ${\cal E} \equiv (q\Omega_{0}\ov{\psi}_{0}/2c)\,\ov{\cal E}$, where
\begin{eqnarray}
\ov{\cal E} &=& \frac{1}{2\chi} \left[ (\chi^{\prime})^{2} \;+\frac{}{} (1 - \chi)^{2} \right] \;+\; \frac{2\epsilon\;\Phi(\ov{\psi}_{0}\chi)}{\ov{\psi}_{0}\,\Phi^{\prime}(\ov{\psi}_{0})} \nonumber \\
 &\equiv& \frac{(\chi^{\prime})^{2}}{2\chi} \;+\; U(\chi),
\label{eq:energy}
\end{eqnarray} 
so that the solution for $\chi(t')$ can be found by inverting the integral solution
\begin{equation}
t'(\chi) \;\equiv\; \Omega_{0}\,t(\chi) \;=\; \pm\int_{\chi_{b}}^{\chi} \frac{ds}{\sqrt{2 s\,[\ov{\cal E} - U(s)]}},
\label{eq:t_chi}
\end{equation}
where the initial condition $\chi_{b}$ can be chosen as a turning point ($\dot{\chi} = 0$) defined by the condition $\ov{\cal E} = U(\chi_{b})$. 

Lastly, the solution for Eq.~\eqref{eq:t_chi} can be analytically obtained by quadrature for an electric potential $\Phi(\psi)$ represented as a polynomial up to third order in $\psi$. Once the solution $\chi(t')$ is found by inverting the integral solution for Eq.~\eqref{eq:t_chi}, the solution for the azimuthal angle $\theta(t')$ is found by integrating
\begin{equation}
\theta^{\prime}(t') \;=\; \frac{1}{2} \left(\frac{1}{\chi(t')} \;-\; 1 \right).
\label{eq:theta_prime}
\end{equation}
Hence, a particle orbit can be generated from these solutions, which can be expressed in polar form as
\begin{eqnarray}
x(t') &=& r(t')\;\cos\theta(t'), \label{eq:x_action} \\
y(t') &=&   r(t')\;\sin\theta(t'), \label{eq:y_action}
\end{eqnarray}
where $r(t') = \sqrt{2\psi(t')/B_{0}}$.

\section{\label{sec:particle_linear}Particle Dynamics in a Linear Radial Electric Field}

In previous work by White, Hassam, and Brizard \cite{White:2018}, a nonuniform radial electric field was represented by the electric potential $\Phi(\psi) = a\,\sqrt{\psi} + b\,\psi$, where $a$ and $b$ are constants associated with a uniform radial electric field and a radial electric with constant gradient, respectively.

Here, we look at charged particle dynamics in the electric potential
\begin{equation}
\Phi(\psi) \;=\; \Phi^{\prime}(\ov{\psi}_{0})\,\psi,
\label{eq:Phi_linear}
\end{equation}
so that, using $\psi = B_{0}\,(x^{2} + y^{2})/2$, we find a linear radial electric field
\begin{equation}
{\bf E}\;=\; -\,\nabla\Phi \;=\; -\,B_{0}\,\Phi^{\prime}(\ov{\psi}_{0}) \left( x\,\wh{\sf x} + y\,\wh{\sf y}\right).
\label{eq:E_linear}
\end{equation}
This nonuniform radial electric field yields a nonuniform $E\times B$ velocity
\begin{equation}
{\bf u} \;=\; {\bf E}\btimes\frac{c\wh{\sf z}}{B_{0}} \;=\; \epsilon\,\Omega_{0}\left(x\,\wh{\sf y} - y\,\wh{\sf x}\right),
\label{eq:ExB_linear}
\end{equation}
which has constant parallel vorticity $\wh{\sf z}\bdot\nabla\btimes{\bf u} = 2\epsilon\,\Omega_{0}$. 

In the ordering recently considered by Joseph \cite{Joseph:2021}, an electric oscillation frequency is defined as $\Omega_{E} \equiv (-q\,\nabla\bdot{\bf E}/mc)^{\frac{1}{2}}$, so that we find $\Omega_{E} = \Omega_{0}\,\sqrt{2\,\epsilon}$ from Eqs.~\eqref{eq:epsilon} and \eqref{eq:E_linear}. Hence, according to the Joseph ordering \cite{Joseph:2021}, our study is situated between the large-flow ordering $\Omega_{E}/\Omega_{0} = {\cal O}(\epsilon)$ and the maximal ordering $\Omega_{E}/\Omega_{0} = {\cal O}(1)$.

In recent work, Kabin \cite{Kabin:2023} considered an additional divergenceless electric field ${\bf E}_{1} = E^{\prime}_{10}\,(y\,\wh{\sf x} + x\,\wh{\sf y})$, but since it is generated by an electric potential $\Phi_{1}(\psi,\theta) = -\,E^{\prime}_{10}\,x\,y = -\,
(E^{\prime}_{10}/B_{0})\,\psi\,\sin(2\theta)$ that breaks the invariance of the azimuthal canonical angular momentum, it will not be considered here.

\subsection{Normal-mode analysis}

Using the electric field \eqref{eq:E_linear}, the equations of motion are expressed in Cartesian coordinates as
\begin{eqnarray}
x^{\prime\prime} &=& -\,\epsilon\,x \;+\; y^{\prime}, \label{eq:x_pp} \\
y^{\prime\prime} &=& -\,\epsilon\,y \;-\; x^{\prime}, \label{eq:y_pp}
\end{eqnarray} 
which have an azimuthal canonical angular momentum invariant
\begin{equation}
\ov{p}_{\theta} \;=\; x\,y^{\prime} \;-\;y\,x^{\prime} \;+\; \frac{1}{2}\,|{\bf x}|^{2} \;\equiv\; 1,
\label{eq:p_linear}
\end{equation}
which follows from Eq.~\eqref{eq:p_theta0}, and an energy invariant
\begin{eqnarray}
\ov{\cal E} &=& |{\bf x}^{\prime}|^{2} \;+\; \epsilon\,|{\bf x}|^{2} \;+\; \frac{1}{|{\bf x}|^{2}}\left(1 \;-\frac{}{} \ov{p}_{\theta}^{2}\right) \;+\; \ov{p}_{\theta} \;-\; 1 \nonumber \\
 &\equiv& |{\bf x}^{\prime}|^{2} \;+\; \epsilon\,|{\bf x}|^{2},
\label{eq:energy_linear}
\end{eqnarray}
which follows from Eqs.~\eqref{eq:energy} and \eqref{eq:p_linear}. We note that, since the equations \eqref{eq:x_pp}-\eqref{eq:y_pp} are linear in $x$ and $y$, they can be arbitrarily normalized.

Using the standard normal-mode analysis, where $x = \ov{x}\,\exp(i\omega t^{\prime})$ and $y = \ov{y}\,\exp(i\omega t^{\prime})$, we obtain the matrix equation
\begin{equation}
\left( \begin{array}{cc}
\epsilon - \omega^{2} & -\,i\,\omega \\
i\,\omega & \epsilon - \omega^{2}
\end{array}\right) \cdot \left(\begin{array}{c}
\ov{x} \\
\ov{y}
\end{array} \right) \;=\; 0,
\end{equation}
which has non-trivial solutions only if
\begin{equation}
\left(\epsilon - \omega^{2}\right)^{2} \;-\; \omega^{2} \;=\; 0,
\end{equation}
with solutions $\pm\,\omega_{+}$ and $\pm\,\omega_{-}$, where
\begin{equation}
\omega_{\pm} \;=\; \frac{1}{2}\left( 1 \;\pm\frac{}{} \sqrt{1 + 4\,\epsilon}\right),
\label{eq:omega_pm}
\end{equation}
and $\omega_{-} < 0 < \omega_{+}$ under the assumption that $\epsilon > 0$. By inspection, the general solutions for Eqs.~\eqref{eq:x_pp}-\eqref{eq:y_pp} are
\begin{eqnarray}
x(t') &=& a\,\cos(\omega_{+}t' + \alpha) \;+\; b\,\cos(\omega_{-}t' + \beta), \label{eq:x_sol} \\
y(t') &=& -\,a\,\sin(\omega_{+}t' + \alpha) \;-\; b\,\sin(\omega_{-}t' + \beta), \label{eq:y_sol}
\end{eqnarray} 
where the constants $(a,\alpha; b,\beta)$ are chosen from initial conditions. We note that the normalized magnetic flux $\chi \equiv (x^{2} + y^{2})/2$ is expressed as
\begin{equation}
\chi(t') \;=\;  \frac{1}{2} \left( a^{2} \;+\frac{}{} b^{2} \right) \;+\; a\,b\;\cos(\tau + \delta),
 \label{eq:chi_cartesian}
 \end{equation}
where $\tau \equiv (\omega_{+} - \omega_{-})t^{\prime} = \sqrt{1 + 4 \epsilon}\,t^{\prime}$ and $\delta \equiv \alpha - \beta$.

\subsection{Integral orbital solution}

We now consider the integral orbital solution \eqref{eq:t_chi}, where the effective potential
\begin{equation}
U(\chi) \;=\; \frac{(1 - \chi)^{2}}{2\chi} \;+\; 2\epsilon\,\chi
\label{eq:U_linear}
\end{equation}
has a minimum $U(\chi_{0}) = 1/\chi_{0} - 1$ at 
\begin{equation}
\chi_{0} \;\equiv\; 1/\sqrt{1 + 4 \epsilon}. 
\label{eq:chi_0}
\end{equation}
Hence, a real orbital radial solution exists for $\ov{\cal E} \geq U(\chi_{0})$ and the radicand in Eq.~\eqref{eq:t_chi} can be expressed as
\[ 2s\,[\ov{\cal E} - U(s)] \;=\; (1 + 4\epsilon)\left[ \tan^{2}\phi \;-\frac{}{} (s - \sec\phi)^{2} \right], \] 
where we defined 
\begin{equation}
(1 + \ov{\cal E})\chi_{0} \;\equiv\; \sec\phi \;\geq\; 1,
\label{eq:energy_sec}
\end{equation}
i.e., the radial motion is periodic when the energy is above the minimum of $U(\chi)$, with $0 \leq \phi < \pi/2$. The orbital solution is, therefore, expressed as
\begin{equation}
\chi(t') \;\equiv\; \chi_{0}\left(\sec\phi \;-\frac{}{} \tan\phi\;\cos\tau\right),
\label{eq:chi_orbit}
\end{equation}
where $\tau = t'/\chi_{0} = \sqrt{1 + 4 \epsilon}\,t^{\prime}$ and $\chi(0)$ is chosen to be at the lower turning point: $\chi(0) = \chi_{0}\,(\sec\phi - \tan\phi)$. By comparing this solution with the normal-mode solution \eqref{eq:chi_cartesian}, we obtain $\delta = \pi$, with $a^{2} + b^{2} = 2\chi_{0}\,\sec\phi$ and $a\,b = \chi_{0}\,\tan\phi$, from which we obtain
\begin{eqnarray}
a(\epsilon,\phi) &=& b(\epsilon,\phi)\,\tan(\phi/2), \nonumber \\
 && \label{eq:ab_phi} \\
 b(\epsilon,\phi) &=& \sqrt{\chi_{0}(\epsilon)\,(1 + \sec\phi)}, \nonumber
 \end{eqnarray}
so that Eqs.~\eqref{eq:energy_linear}-\eqref{eq:p_linear} yield
\begin{eqnarray*}
\ov{p}_{\theta} &=& a^{2} \left(\frac{1}{2} - \omega_{+}\right) \;+\; b^{2} \left(\frac{1}{2} - \omega_{-}\right) \;=\; 1, \\
\ov{\cal E} &=& a^{2}\,\left(\omega_{+}^{2} \;+\frac{}{} \epsilon\right) \;+\; b^{2}\,\left(\omega_{-}^{2} \;+\frac{}{} \epsilon\right) \;=\; \sec\phi/\chi_{0} - 1,
\end{eqnarray*}
which follow from Eqs.~\eqref{eq:p_theta0} and \eqref{eq:energy_sec}, respectively.

Using the orbital solution \eqref{eq:chi_orbit}, the solution for the azimuthal angle is obtain from Eq.~\eqref{eq:theta_prime} as
\begin{eqnarray}
\theta(t') &=& -\,\frac{t'}{2} \;+\; \frac{1}{2} \int_{0}^{\tau} \frac{dz}{\sec\phi - \tan\phi\;\cos z} \nonumber \\
 &\equiv& -\,\frac{t'}{2} \;+\; \vartheta(\tau,\phi),
 \label{eq:theta_orbit}
\end{eqnarray} 
where 
\begin{eqnarray}
\vartheta(\tau,\phi) &=& \arctan\left[i \left( \sec\phi \;-\frac{}{} \tan\phi\;e^{i\tau}\right)\right] \nonumber \\
 &&-\;  \arctan\left[i \left( \sec\phi \;-\frac{}{} \tan\phi\right)\right] \nonumber \\
  &\equiv& -\,\frac{i}{2} \ln\left( \frac{1 - e^{i\tau}\,\cot(\phi/2)}{e^{i\tau} - \cot(\phi/2)}\right),
\end{eqnarray}
which vanishes at $\tau = 0$ and, as expected from Eq.~\eqref{eq:theta_orbit}, $\vartheta(\tau,\phi) \rightarrow \tau/2$ as $\phi \rightarrow 0$, i.e., $\theta(t') \rightarrow -\,\omega_{-}\,t'$.

From the radial solution \eqref{eq:chi_orbit}, we find that the radial period is $T = 2\pi\,\chi_{0}$ and the azimuthal angular deviation between successive radial maxima (or minima) is obtained from Eq.~\eqref{eq:theta_orbit} as $\Delta\theta \equiv \theta(T) - \theta(0) = \pi \,(1 - \chi_{0})$, which implies that the planar curve $(x(t'),y(t'))$ closes upon itself only if $\chi_{0}$ is a rational number. We also note that the planar curve initiates retrograde motion near the upper radial turning point $\chi_{0}\,(\sec\phi + \tan\phi)$ when $\phi > \arcsin[(1 - \chi_{0}^{2})/(1 + \chi_{0}^{2})] = \arcsin[2\epsilon/(1 + 2\epsilon)]$.

 \begin{figure}
\epsfysize=2.8in
\epsfbox{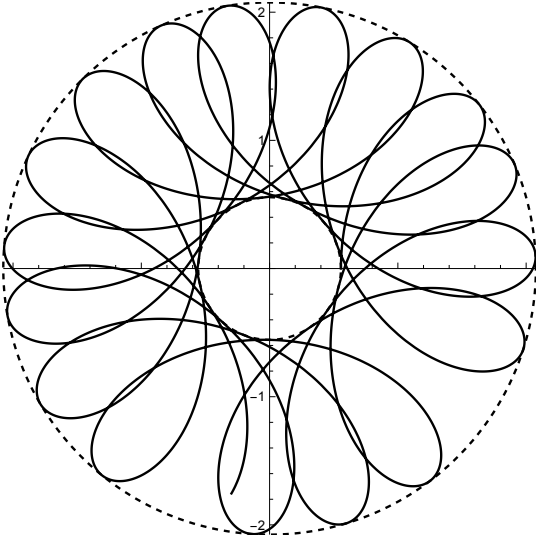}
\caption{Plot of the planar curve $(x(t'),y(t'))$, given by Eqs.~\eqref{eq:x_final}-\eqref{eq:y_final}, for $\epsilon = 1/2$ ($\chi_{0} = 1/\sqrt{3}$) and $\phi = \pi/3$. The outer and inner circles, with radii $\sqrt{2\chi_{0}\,(\sec\phi \pm \tan\phi)} = \sqrt{(4/\sqrt{3}) \pm 2}$ are shown as dashed circles.}
\label{fig:plotxy}
\end{figure}

 \begin{figure}
\epsfysize=2.8in
\epsfbox{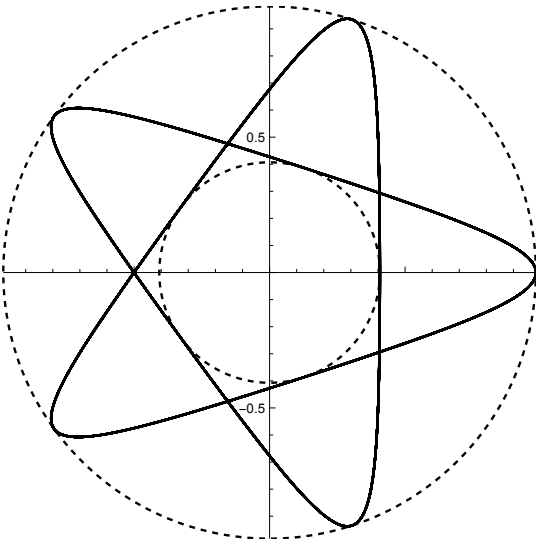}
\caption{Plot of the planar curve $(x(t'),y(t'))$, given by Eqs.~\eqref{eq:x_final}-\eqref{eq:y_final}, for $\epsilon = 6$ ($\chi_{0} = 1/5$) and $\phi = \pi/4$. The outer and inner circles, with radii $\sqrt{2\chi_{0}\,(\sec\phi \pm \tan\phi)} = \sqrt{2\,(\sqrt{2} \pm 1)/5}$ are shown as dashed circles.}
\label{fig:plotxy_2}
\end{figure}

Figures \ref{fig:plotxy}-\ref{fig:plotxy_2} show two cases parametrized by different values of $(\epsilon,\phi)$. In Fig.~\ref{fig:plotxy}, the value $\epsilon = 1/2$ causes $\chi_{0} = 1/\sqrt{3}$ to be irrational and the planar curve $(x(t'),y(t'))$ does not close upon itself. The planar curve also exhibits retrograde motion since $\phi = \pi/3 >  \arcsin(1/2) = \pi/6$. In Fig.~\ref{fig:plotxy_2}, on the other hand, the value $\epsilon = 6$ is chosen so that $\chi_{0} = 1/5$ is rational and, therefore, the planar curve 
$(x(t'),y(t'))$ closes upon itself (after 5 radial cycles). Since $\phi = \pi/4 <  \arcsin(12/13) \simeq 3\pi/8$, however, the planar curve does not exhibit retrograde motion.

When expressed in terms of Cartesian coordinates, the orbital solution is expressed as
\begin{eqnarray}
x(t') &=& \sqrt{2\,\chi(t')}\;\cos\theta(t') \label{eq:x_final} \\
 &=& b(\epsilon,\phi) \left[ \cos(\omega_{-}t') \;-\frac{}{} \tan(\phi/2)\;\cos(\omega_{+}t')\right], \nonumber \\
y(t') &=& \sqrt{2\,\chi(t')}\;\sin\theta(t') \label{eq:y_final} \\
 &=& b(\epsilon,\phi) \left[ \tan(\phi/2)\;\sin(\omega_{+}t') \;-\frac{}{} \sin(\omega_{-}t')\right], \nonumber
\end{eqnarray}
where we selected the phases $\alpha = \pi$ and $\beta = 0$. We note that, when the orbital solution \eqref{eq:chi_orbit}-\eqref{eq:theta_orbit} is evaluated at the minimum $(\phi = 0)$ of the effective potential $U(\chi)$, we find a circular solution, with a constant radius $b = \sqrt{2\chi_{0}}$ (with $a = 0$), and $\theta(t') = -\,\omega_{-}\,t^{\prime}$.

This completes our analysis of the charged-particle motion in a uniform magnetic field $B_{0}\,\wh{\sf z} = \nabla\psi\btimes\nabla\theta$ with a linear radial electric field ${\bf E} = -\,\nabla\Phi = -\,\Phi^{\prime}_{0}\,\nabla\psi$ with constant $E\times B$ parallel vorticity.

\section{\label{sec:nonlinear_gc}Guiding-center Analysis for a Nonlinear Radial Electric Field}

In this Section, we proceed with the guiding-center analysis of a general radial electric field ${\bf E} = -\,\nabla\Phi(\psi)$ with the dimensionless parameter \eqref{eq:epsilon} considered in the limit $\epsilon \ll 1$. The purpose of the guiding-center analysis is to derive a reduced dynamical description in which the fast gyromotion has been transformed away (not averaged!). 

The Hamiltonian guiding-center theory of charged-particle motion in the presence of electric and magnetic fields was presented in Refs.~\cite{Brizard:1995, Cary_Brizard:2009}, and was recently summarized in Ref.~\cite{Frei_etal:2020}, for the case of a nonuniform magnetic field. Here, we apply the same perturbation analysis for the simpler case of a uniform magnetic field.

\subsection{\label{sec:ordering}Particle Lagrangian in a drifting frame}

The guiding-center analysis begins by shifting the lab reference frame to a reference frame drifting with the $E\times B$ velocity
\begin{eqnarray}
{\bf u} &=& \frac{c\wh{\sf z}}{B_{0}}\btimes\nabla\Phi \;=\; c\,\Phi^{\prime}(\psi)\;\pd{\bf x}{\theta} \;=\; \epsilon\,\Omega_{0}\,\nu(\psi) \;\pd{\bf x}{\theta} \nonumber \\
 &=&  \frac{2c}{B_{0}}\,\Phi^{\prime}(\psi)\,\psi\;\nabla\theta \;=\; \frac{\Omega_{0}}{B_{0}}\;\epsilon\,\Psi_{1}(\psi) \;\nabla\theta,
\label{eq:ExB}
\end{eqnarray}
which is directed along the azimuthal direction, with parallel $E\times B$ parallel vorticity 
\begin{equation}
\wh{\sf z}\bdot\nabla\btimes{\bf u} \;=\; \frac{c}{B_{0}}\,\nabla^{2}\Phi \;=\; \epsilon\,\Omega_{0}\,\Psi_{1}^{\prime}(\psi). 
\label{eq:vorticity}
\end{equation}
Here, the first-order correction $\Psi_{1}(\psi)$ is defined as
\begin{equation}
\Psi_{1}(\psi) \;\equiv\; \psi \left(2\,\frac{c\Phi^{\prime}(\psi)}{\epsilon\,\Omega_{0}}\right) \;=\; 2\;\psi\,\nu(\psi),
\label{eq:Psi_1}
\end{equation}
so that the phase-space position of a charged particle is transformed as $({\bf x},{\bf v}) \rightarrow ({\bf x},{\bf w})$, where ${\bf w} \equiv {\bf v} - {\bf u}$ denotes the relative particle velocity in the drifting frame. 

Hence, the shifted particle Lagrangian becomes
\begin{eqnarray}
L_{E} &=& \left[\frac{q}{c}\,\psi\,\nabla\theta \;+\; m\,({\bf w} + {\bf u})\right]\bdot\dot{\bf x} \nonumber \\
 &&-\; \left( q\,\Phi \;+\; \frac{m}{2}|{\bf w} + {\bf u}|^{2} \right),
\label{eq:L_E}
\end{eqnarray}
where $|{\bf w} + {\bf u}|^{2} = |{\bf w}|^{2} + |{\bf u}|^{2} + 2\,{\bf w}\bdot{\bf u}$. We note that we restrict our analysis to two-dimensional motion in the $(x,y)$-plane, where ${\bf w}$ is expressed in particle space as
\begin{equation}
{\bf w} \;=\; \dot{\bf x} \;-\; {\bf u} \;=\; \dot{\psi}\,\pd{\bf x}{\psi} \;+\; \left[ \dot{\theta} \;-\frac{}{} c\,\Phi^{\prime}(\psi)\right]\pd{\bf x}{\theta}.
\label{eq:w_def}
\end{equation}
Here, the magnitude of ${\bf w}$ depends on the lowest-order magnetic moment $\mu_{0}$: $w = |{\bf w}| = \sqrt{2\,\mu_{0}B_{0}/m}$, while the unit vector $\wh{\bot} \equiv {\bf w}/w$ depends on the spatial coordinates $(\psi,\theta)$ as well as the lowest-order gyroangle $\zeta_{0}$.

We now consider a guiding-center transformation $(\psi,\theta; {\bf w}) \rightarrow (\Psi,\Theta; \mu,\zeta)$, where $(\Psi,\Theta)$ denote the guiding-center coordinates, $\mu$ denotes the guiding-center magnetic moment, and $\zeta$ denotes the guiding-center gyroangle that is canonically conjugate to the guiding-center gyroaction $J = \mu\,B_{0}/\Omega_{0}$. The analysis begins with renormalizing the mass of the particle as $m \rightarrow \epsilon\,m$ (which is analogous to performing an expansion in $1/\Omega_{0}$), so that the shifted particle Lagrangian is expressed as $L_{E} = L_{E0} + \epsilon\,L_{E1}$, where the lowest-order particle Lagrangian is 
\begin{equation}
L_{E0} \;=\; (q/c)\psi\,\dot{\theta} \;-\; q\,\Phi(\psi), 
\label{eq:LE_0}
\end{equation}
while the first-order particle Lagrangian is
\begin{eqnarray}
L_{E1} &=& m\,\left[ {\bf w}\bdot\left( \pd{\bf x}{\psi}\,\dot{\psi} +  \pd{\bf x}{\theta}\,\dot{\theta} \right) \;-\; {\bf w}\bdot c\,\Phi^{\prime}(\psi)\pd{\bf x}{\theta} \right] \nonumber \\
 &&+\; \frac{q}{c}\,\Psi_{1}(\psi)\,\dot{\theta} \;-\; \left( q\,\Phi_{1}(\psi) \;+\frac{}{} \mu_{0}\,B_{0} \right),
 \label{eq:LE_1}
\end{eqnarray}
which explicitly displays the gyroangle-dependent relative velocity ${\bf w}$, and 
\begin{equation}
\frac{1}{2}m|{\bf u}|^{2} \;=\; q\,\psi\,\Phi^{\prime}\;\left(\frac{c\Phi^{\prime}}{\Omega_{0}}\right) \;\equiv\; \epsilon\,q\,\Phi_{1}(\psi)
\label{eq:Phi_1}
\end{equation}
introduces the first-order correction $\Phi_{1}(\psi)$ to the electrostatic potential $\Phi(\psi)$. We note that the gradient of $\Phi_{1}$ introduces centrifugal effects in the guiding-center dynamics of a charged particle \cite{Peeters:2009}.

\subsection{Guiding-center dynamics in a drifting frame}

The purpose of the guiding-center transformation is to remove the linear contributions from the gyroangle-dependent relative velocity ${\bf w}$ from Eq.~\eqref{eq:LE_1}. As a result of this transformation, the shifted guiding-center Lagrangian is generically expressed as
\begin{equation}
L_{E{\rm gc}} \;\equiv\; \frac{q}{c}\,\Psi^{*}\dot{\Theta} \;+\; J\left(\dot{\zeta} \;-\frac{}{} \dot{\Theta}\right) - \left( q\,\Phi^{*} + \mu\;B^{*}\right),
\label{eq:LE_gc}
\end{equation}
where $(\Psi^{*},\Phi^{*},B^{*})$ are functions of $\Psi$ that will be derived after the guiding-center transformation is defined (see Sec.~\ref{sec:gc_Lag}). We note that the terms $J\,(\dot{\zeta} - \dot{\Theta})$ appear in order to satisfy gyrogauge invariance, with the gyrogauge vector $\vb{\cal R} \equiv \nabla\wh{\sf e}_{1}\bdot\wh{\sf e}_{2} = \nabla\Theta$ calculated from cylindrical geometry, so that $\vb{\cal R}\bdot\dot{\bf X} = \dot{\Theta}$.

The guiding-center equation of motion for the two-dimensional guiding-center position ${\bf X}$ is obtained from the guiding-center Lagrangian \eqref{eq:LE_gc} as
\begin{equation}
\dot{\bf X} \;=\; \frac{c\wh{\sf z}}{qB_{\|}^{*}}\btimes\left(q\,\nabla\Phi^{*} \;+\; \mu\;\nabla B^{*}\right),
\label{eq:Xgc_dot}
\end{equation}
where 
\begin{equation}
B_{\|}^{*} \equiv \wh{\sf z}\bdot{\bf B}^{*} \;=\;  \wh{\sf z}\bdot\nabla\Psi^{*}\btimes\nabla\Theta \;=\; B_{0}\;d\Psi^{*}/d\Psi,
\end{equation}
while the equation for the guiding-center gyroangle $\zeta$ is expressed as
\begin{equation}
\dot{\zeta} \;=\; \Omega_{0}\;B^{*}/B_{0} \;+\; \vb{\cal R}\bdot\dot{\bf X}.
\end{equation}
From Noether's Theorem \cite{Brizard:2015}, we easily conclude that $\Psi$ and $\mu$ are guiding-center constants of motion since the guiding-center Lagrangian \eqref{eq:LE_gc} is independent of the angles $\Theta$ and $\zeta$.

When considering the guiding-center motion in physical space, we find the Cartesian representation for a circle: $X(t) = R\,\cos\Theta(t)$ and $Y(t) = R\,\sin\Theta(t)$, with radius $R \equiv \sqrt{2\Psi/B_{0}}$. We also immediately find that the guiding-center energy 
\begin{equation}
{\cal E}_{\rm gc} \;=\; q\,\Phi^{*}(\Psi) \;+\; \mu\;B^{*}(\Psi) 
\label{eq:E_gc} 
\end{equation}
and the guiding-center azimuthal canonical angular momentum 
\begin{equation}
P_{\Theta{\rm gc}} \;\equiv\; \partial L_{E{\rm gc}}/\partial\dot{\Theta} \;=\; (q/c)\,\Psi^{*}(\Psi) - J 
\label{eq:P_gc}
\end{equation}
are guiding-center constants of motion.

We will now construct explicit expressions for $(\Psi^{*},\Phi^{*},B^{*})$ as functions of $\Psi$, represented as expansions in powers of $\epsilon$, once again interpreted through the mass renormalization $m \rightarrow \epsilon\,m$.

\subsection{Guiding-center transformation}

The derivation of the guiding-center transformation that leads from the particle Lagrangian \eqref{eq:L_E} to the guiding-center Lagrangian \eqref{eq:LE_gc} begins with the separation of a generic Lagrangian $L = p_{\alpha}\,\dot{z}^{\alpha} - H$ into a symplectic part $p_{\alpha}\dot{z}^{\alpha}$, which is then converted into the symplectic one-form $\gamma = p_{\alpha}\,\exd z^{\alpha}$ (where $\exd$ denotes an exterior derivative), and a Hamiltonian part $H$. 

Next, we construct the guiding-center transformation as an asymptotic expansion in powers of $\epsilon$ for each guiding-center phase-space coordinate $Z^{\alpha} = (\Psi,\Theta,\mu,\zeta)$ in terms of the particle phase-space coordinates $z^{\alpha} = (\psi,\theta,\mu_{0},\zeta_{0})$: 
\begin{equation}
Z^{\alpha} \;=\; z^{\alpha} \;+\; \epsilon\,G_{1}^{\alpha} \;+\; \epsilon^{2} \left( G_{2}^{\alpha} \;+\; \frac{1}{2}\;G_{1}^{\beta}\pd{G_{1}^{\alpha}}{z^{\beta}} \right) + \cdots,
\label{eq:Zz_gc}
\end{equation}
where the components $(G_{n}^{\psi},G_{n}^{\theta},G_{n}^{\mu},G_{n}^{\zeta})$ are chosen at $n$th-order in order to derive an $n$th-order guiding-center Lagrangian that is independent of the guiding-center gyroangle. Once these components are derived, we return the particle mass to its physical value $\epsilon m \rightarrow m$. 

Using the standard methods of Lie-transform perturbation theory \cite{Littlejohn:1982}, the new symplectic one-form 
\begin{equation}
{\sf T}_{\rm gc}^{-1}\gamma + \exd S \;\equiv\; P_{\alpha}(\Psi,\Theta;\mu)\,\exd Z^{\alpha}
\end{equation}
where $S$ is an arbitrary gauge function, and the new Hamiltonian 
 \begin{equation}
 {\sf T}_{\rm gc}^{-1}H \;\equiv\;  H_{\rm gc}(\Psi,\Theta;\mu)
 \end{equation}
 are obtained at each order in $\epsilon$, where the guiding-center push-forward operator 
 \[ {\sf T}_{\rm gc}^{-1} \;=\; \cdots \exp(-\epsilon^{2}\pounds_{2})\,\exp(-\epsilon\pounds_{1}) \]
 is expressed in terms of Lie derivatives $\pounds_{n}$ generated by the vector field ${\sf G}_{n}$, which are then used in the guiding-center transformation \eqref{eq:Zz_gc}.

Using the ordering \eqref{eq:epsilon}, the phase-space Lagrangian symplectic one-form $\gamma_{E} = \gamma_{E0} + \epsilon\,\gamma_{E1}$ is expressed as 
\begin{eqnarray}
\gamma_{E0} &=& (q/c)\,\psi\,\exd\theta, \label{eq:gamma_0} \\
\gamma_{E1} &=&\frac{q}{c}\,\Psi_{1}(\psi)\;\exd\theta \;+\; m{\bf w}\bdot\left(\pd{\bf x}{\psi}\,\exd\psi + \pd{\bf x}{\theta}\,\exd\theta\right), \label{eq:gamma_1}
\end{eqnarray}
where $\Psi_{1}(\psi)$ is defined in Eq.~\eqref{eq:Psi_1}, while the zeroth and first-order Hamiltonians, on the other hand, are
\begin{eqnarray}
H_{E0} &=& q\,\Phi(\psi), \label{eq:Ham_0} \\
H_{E1} &=& q\,\Phi_{1}(\psi) \;+\; \mu_{0}\,B_{0} \;+\; m\,{\bf w}\bdot c\,\Phi^{\prime}(\psi)\pd{\bf x}{\theta}, \label{eq:Ham_1}
\end{eqnarray}
where $\Phi_{1}(\psi)$ is defined in Eq.~\eqref{eq:Phi_1}.

\subsubsection{Zeroth-order analysis}

By definition, the zeroth-order guiding-center symplectic one-form is 
\begin{equation}
\Gamma_{E{\rm gc}0} \;\equiv\; (q/c)\,\Psi\,\exd\Theta,
\label{eq:Gamma_0}
\end{equation}
where $(\Psi,\Theta)$ denotes the guiding-center position. The zeroth-order guiding-center Hamiltonian, on the other hand, is
\begin{equation}
H_{E{\rm gc}0} \;\equiv\; q\,\Phi(\Psi),
\label{eq:H_0}
\end{equation}
so that the zeroth-order guiding-center Lagrangian is
\begin{equation}
L_{E{\rm gc}0} \;=\; (q/c)\Psi\,\dot{\Theta} \;-\; q\,\Phi(\Psi), 
\label{eq:LEgc_0}
\end{equation}
which yields the zeroth-order equation of motion $\dot{\Theta} = c\,\Phi^{\prime}(\Psi)$ and the azimuthal canonical angular momentum conservation law is $(q/c)\,\dot{\Psi} = \partial L_{E{\rm gc}0}/\partial\Theta = 0$ implies that $\Psi$ is conserved at the lowest order. 

\vspace*{0.2in}
\subsubsection{First-order analysis}

Next, the first-order guiding-center symplectic one-form is constructed as
\begin{eqnarray}
\Gamma_{E{\rm gc}1} &=& \frac{q}{c}\Psi_{1}\;\exd\theta \;+\; m{\bf w}\bdot\left(\pd{\bf x}{\psi}\,\exd\psi + \pd{\bf x}{\theta}\,\exd\theta\right) \nonumber \\
 &&-\; \frac{q}{c} \left(G_{1}^{\psi}\,\exd\theta \;-\frac{}{} G_{1}^{\theta}\,\exd\psi\right) \nonumber \\
  &\equiv&  \frac{q}{c}\Psi_{1}(\Psi)\;\exd\Theta,
 \label{eq:Gamma_1}
\end{eqnarray}
where $S_{1} = 0$ at this order, and the gyroangle-dependent relative velocity ${\bf w}$ is removed by choosing the spatial components
\begin{eqnarray}
G_{1}^{\psi} &=& (B_{0}/\Omega_{0})\,{\bf w}\bdot\partial{\bf x}/\partial\theta, \label{eq:G1_psi} \\
G_{1}^{\theta} &=& -\,(B_{0}/\Omega_{0})\,{\bf w}\bdot\partial{\bf x}/\partial\psi, \label{eq:G1_theta}
\end{eqnarray}
which yields the standard result \cite{Cary_Brizard:2009}
\begin{equation}
G_{1}^{\bf x} \;=\; {\bf w}\btimes\frac{\wh{\sf z}}{\Omega_{0}} \;=\; \frac{1}{\Omega_{0}}\,\pd{\bf w}{\zeta} \;\equiv\; -\;\vb{\rho},
\label{eq:G1_x}
\end{equation}
where the relative velocity ${\bf w} \equiv \Omega_{0}\,\partial\vb{\rho}/\partial\zeta_{0}$ is defined in Eq.~\eqref{eq:w_def}. We note that by returning the particle mass to its physical value $\epsilon m \rightarrow m$, the components \eqref{eq:G1_psi}-\eqref{eq:G1_theta} are, in fact, zeroth-order in $\epsilon$ and, therefore, we will need to derive the components $(G_{2}^{\psi},G_{2}^{\theta})$ at second order.

The first-order guiding-center Hamiltonian is constructed as
\begin{eqnarray}
H_{E{\rm gc}1} &=& q\,\Phi_{1} \;+\; \mu_{0}\,B_{0} \;+\;  \frac{B_{0}}{\Omega_{0}}\,q\,\Phi^{\prime} \;{\bf w}\bdot\pd{\bf x}{\theta} \;-\; q\,\Phi^{\prime}\;G_{1}^{\psi} \nonumber \\
 &=&  q\,\Phi_{1}(\Psi) \;+\; \mu\,B_{0},
 \label{eq:H_1}
\end{eqnarray}
where we used Eq.~\eqref{eq:G1_psi} to cancel the gyroangle-dependent relative velocity ${\bf w}$. Hence, the first-order guiding-center Lagrangian is
\begin{equation}
L_{E{\rm gc}1} \;=\; \frac{q}{c}\,\Psi_{1}(\Psi)\,\dot{\Theta} \;-\; \left(q\,\Phi_{1}(\Psi) \;+\frac{}{} \mu\,B_{0}\right), 
\label{eq:LEgc_1}
\end{equation}
which preserves the conservation law of $\Psi$ of the zeroth-order guiding-center Lagrangian.

\subsubsection{Second-order analysis}

At second order, the second-order guiding-center symplectic one-form is constructed as
\begin{widetext}
\begin{eqnarray}
\Gamma_{E{\rm gc}2} &=& -\; \frac{q}{c} \left[ \left(G_{2}^{\psi}\,\exd\theta - G_{2}^{\theta}\,\exd\psi\right) \;+\frac{}{} \Psi_{1}^{\prime}  \left(G_{1}^{\psi}\,\exd\theta - G_{1}^{\theta}\,\exd\psi\right)\right] \;-\; \frac{m}{2} \left( G_{1}^{\mu}\,\pd{\bf w}{\mu_{0}} + G_{1}^{\zeta}\,\pd{\bf w}{\zeta_{0}}\right)\bdot\left( \pd{\bf x}{\psi}\,\exd\psi + \pd{\bf x}{\theta}\,\exd\theta\right) \nonumber \\
  &&+\; \frac{m}{2}\;G_{1}^{\bf x}\bdot\left[ \left( \pd{\bf w}{\mu_{0}}\,\exd\mu_{0} + \pd{\bf w}{\zeta_{0}}\,\exd\zeta_{0}\right) \;-\; \exd{\bf x}\btimes\nabla\btimes{\bf w} \right] \;\equiv\; J\,\left(\exd\zeta \;-\frac{}{} \vb{\cal R}\bdot\exd{\bf X}\right) \;=\; J\,\left(\exd\zeta \;-\frac{}{} \exd\Theta\right), \label{eq:Gamma_2}
\end{eqnarray}
\end{widetext}
where $S_{2} = 0$ at this order, $J \equiv \mu\,B_{0}/\Omega_{0}$ is the guiding-center gyroaction, with its canonically-conjugate guiding-center gyroangle $\zeta$, and the gyrogauge vector $\vb{\cal R} \equiv \nabla\wh{\sf e}_{1}\bdot\wh{\sf e}_{2} = \nabla\Theta$ is calculated from cylindrical geometry (with $\wh{\sf e}_{1} = \wh{r}$ and $\wh{\sf e}_{2} = \wh{\theta} = \wh{\sf z}\btimes\wh{\sf e}_{1}$). Here, we use the identity
\begin{equation}
\nabla\btimes{\bf w} \;=\; \pd{\bf w}{\zeta_{0}}\btimes\vb{\cal R},
\end{equation}
which follows from the alternate definition $\vb{\cal R} = \nabla\wh{\bot}\bdot\wh{\rho}$, where ${\bf w} \equiv w\,\wh{\bot}$ ($w$ is constant in a uniform magnetic field) and $\wh{\sf z} = \wh{\bot}\btimes\wh{\rho}$, so that
\[ \frac{m}{2}\left(\vb{\rho}\btimes\nabla\btimes{\bf w} \right)\bdot\exd{\bf x} = J_{0}\,\vb{\cal R}\bdot\exd{\bf x} + \frac{m}{2}\,(\vb{\rho}\bdot\vb{\cal R})\pd{\bf w}{\zeta}\bdot\exd{\bf x}. \]
Hence, Eq.~\eqref{eq:Gamma_2} yields the second-order spatial components
\begin{eqnarray}
G_{2}^{\psi} &=& -\,\Psi_{1}^{\prime}\;G_{1}^{\psi} - \frac{B_{0}}{2\Omega_{0}}\left( G_{1}^{\mu}\,\pd{\bf w}{\mu_{0}} + g_{1}^{\zeta}\pd{\bf w}{\zeta_{0}}\right)\vb{\cdot}\pd{\bf x}{\theta}, \label{eq:G2_psi} \\
G_{2}^{\theta} &=& -\,\Psi_{1}^{\prime}\;G_{1}^{\theta} + \frac{B_{0}}{2\Omega_{0}}\left( G_{1}^{\mu}\,\pd{\bf w}{\mu_{0}} + g_{1}^{\zeta}\pd{\bf w}{\zeta_{0}}\right)\vb{\cdot}\pd{\bf x}{\psi},
\end{eqnarray}
where $g_{1}^{\zeta} \equiv G_{1}^{\zeta} + \vb{\rho}\bdot\vb{\cal R}$. The second-order spatial vector field is, therefore, expressed as
\begin{equation}
G_{2}^{\bf x} \;=\; \Psi_{1}^{\prime}\,\vb{\rho} + \frac{1}{2}\left( G_{1}^{\mu}\,\pd{\vb{\rho}}{\mu_{0}} + g_{1}^{\zeta}\pd{\vb{\rho}}{\zeta_{0}}\right),
\label{eq:G2_x}
\end{equation}
where we substituted Eq.~\eqref{eq:G1_x}.

We now turn our attention to the second-order guiding-center Hamiltonian, which is constructed as
\begin{eqnarray}
H_{E{\rm gc}2} &=& -\,q \left( \Phi^{\prime}G_{2}^{\psi} + \Phi_{1}^{\prime}G_{1}^{\psi}\right) - B_{0}\,G_{1}^{\mu} \nonumber \\
 &&+\; \frac{m}{2} \left[ ({\bf u}\vb{\rho}):\nabla{\bf w} \;+\frac{}{} ({\bf w}\vb{\rho}):\nabla{\bf u} \right] \nonumber \\
  &&-\; \frac{m}{2} \left( G_{1}^{\mu}\pd{\bf w}{\mu_{0}} + G_{1}^{\zeta}\pd{\bf w}{\zeta_{0}}\right)\bdot{\bf u}.
  \label{eq:H2_first}
\end{eqnarray}
First, we note that
\begin{equation}  
\frac{m}{2}\;({\bf u}\vb{\rho}):\nabla{\bf w} \;=\; -\,\frac{m{\bf u}}{2}\bdot\pd{\bf w}{\zeta_{0}}\;(\vb{\rho}\bdot\vb{\cal R}),
\end{equation}
while
\begin{equation} 
\frac{m}{2}\;({\bf w}\vb{\rho}):\nabla{\bf u} \;=\; -\,J_{0}{\sf a}_{1}:\nabla{\bf u} \;-\; \frac{1}{2}\,J_{0}\wh{\sf z}\bdot\nabla\btimes{\bf u},
\label{eq:a1_eq}
\end{equation}
where the dyadic tensor ${\sf a}_{1} \equiv -\frac{1}{2}\,(\wh{\bot}\wh{\rho} + \wh{\rho}\wh{\bot})$ is explicitly gyroangle-dependent. Hence, inserting these expressions into Eq.~\eqref{eq:H2_first}, while using Eq.~\eqref{eq:G2_psi}, we obtain
\begin{eqnarray}
H_{E{\rm gc}2} &=& q \left( \Phi^{\prime}\,\Psi_{1}^{\prime} \;-\frac{}{} \Phi_{1}^{\prime}\right) G_{1}^{\psi} \;-\; B_{0}\,G_{1}^{\mu} \nonumber \\
 &&-\; J_{0} \left( {\sf a}_{1}:\nabla{\bf u} \;+\; \frac{\wh{\sf z}}{2}\bdot\nabla\btimes{\bf u} \right),
 \end{eqnarray}
 where the terms $G_{1}^{\psi}$ and ${\sf a}_{1}$ are explicitly gyroangle-dependent and must be removed from the guiding-center Hamiltonian. The second-order guiding-center Hamiltonian is, therefore, defined as
 \begin{equation}
H_{E{\rm gc}2} \;\equiv\; -\,B_{0}\;\left\langle G_{1}^{\mu}\right\rangle \;-\;  J\;\frac{\wh{\sf z}}{2}\bdot\nabla\btimes{\bf u},
\label{eq:H2_final}
\end{equation}
where the gyroangle-dependent part of $G_{1}^{\mu}$ is defined as
\begin{equation}
\wt{G}_{1}^{\mu} \;=\; \frac{q}{B_{0}} \left( \Phi^{\prime}\,\Psi_{1}^{\prime} \;-\frac{}{} \Phi_{1}^{\prime}\right) G_{1}^{\psi} \;-\; \frac{J_{0}}{B_{0}}\;{\sf a}_{1}:\nabla{\bf u}.
\label{eq:G1mu_tilde}
\end{equation}
The remaining first-order components $\langle G_{1}^{\mu}\rangle$ and $G_{1}^{\zeta}$ must now be determined at third order.

By combining the symplectic structure \eqref{eq:Gamma_2} and the Hamiltonian \eqref{eq:H2_final}, the second-order guiding-center Lagrangian is expressed as
\begin{eqnarray}
L_{E{\rm gc}2} &=& J\,\left(\dot{\zeta} - \dot{\Theta}\right) + B_{0}\;\left\langle G_{1}^{\mu}\right\rangle + J\;\frac{\wh{\sf z}}{2}\bdot\nabla\btimes{\bf u},
\end{eqnarray} 
which now introduces the gyromotion dynamics.

\subsubsection{Third-order analysis}

Because of the smallness of the ordering parameter $\epsilon$, there is no interest (at this time) in deriving third-order corrections to the guiding-center Lagrangian. The missing first-order components $(\langle G_{1}^{\mu}\rangle,G_{1}^{\zeta})$, however, are determined at third order in the guiding-center analysis from the identities \cite{Brizard:1995}
\begin{eqnarray}
G_{1}^{\mu} &=& -\; \mu_{0}(\wh{\sf z}/\Omega_{0})\bdot\nabla\btimes{\bf u} \;+\; (\Omega_{0}/B_{0})\;\partial\ov{S}_{3}/\partial\zeta_{0}, \label{eq:G1mu_3} \\
G_{1}^{\zeta} &=& -\;(\Omega_{0}/B_{0})\;\partial S_{3}/\partial\mu_{0},  \label{eq:G1zeta_3}
\end{eqnarray}
where the third-order scalar functions $(S_{3},\ov{S}_{3})$ are explicitly gyroangle-dependent, with
\begin{equation}
\ov{S}_{3} \;\equiv\; S_{3} \;-\; \frac{2}{3}\;\mu_{0}\,(B_{0}/\Omega_{0})\,\vb{\rho}\bdot\vb{\cal R}.
\label{eq:S3_def}
\end{equation}
First, by gyroangle averaging both sides of Eq.~\eqref{eq:G1mu_3}, we immediately find that 
\begin{equation}
\left\langle G_{1}^{\mu}\right\rangle \;\equiv\; -\; \mu\;(\wh{\sf z}/\Omega_{0})\bdot\nabla\btimes{\bf u} \;=\; -\; \mu\;\epsilon\,\Psi_{1}^{\prime},
\label{eq:G1mu_av}
\end{equation}
and the second-order guiding-center Hamiltonian \eqref{eq:H2_final} becomes
\begin{equation}
H_{E{\rm gc}2} \;=\; J\;\frac{\wh{\sf z}}{2}\bdot\nabla\btimes{\bf u} \;=\; \frac{J}{2}\; \epsilon\,\Omega_{0}\,\Psi_{1}^{\prime}.
\label{eq:H2_gc}
\end{equation}
while, using ${\bf w} = \Omega_{0}\,\partial\vb{\rho}/\partial\zeta_{0}$, Eq.~\eqref{eq:G1mu_tilde} yields
\[
\pd{\ov{S}_{3}}{\zeta} \;=\; \frac{qB_{0}}{\Omega_{0}} \left( \Phi^{\prime}\,\Psi_{1}^{\prime} \;-\frac{}{} \Phi_{1}^{\prime}\right) \pd{\vb{\rho}}{\zeta_{0}}\bdot\pd{\bf x}{\theta} \;-\; \frac{J_{0}}{\Omega_{0}}\pd{{\sf a}_{2}}{\zeta_{0}}:\nabla{\bf u},
\]
where ${\sf a}_{1} \equiv \partial{\sf a}_{2}/\partial\zeta$ and ${\sf a}_{2} \equiv \frac{1}{4} (\wh{\bot}\wh{\bot} - \wh{\rho}\wh{\rho})$, which is solved as
\[ \ov{S}_{3} \;=\; \frac{qB_{0}}{\Omega_{0}} \left( \Phi^{\prime}\,\Psi_{1}^{\prime} \;-\frac{}{} \Phi_{1}^{\prime}\right) \vb{\rho}\bdot\pd{\bf x}{\theta} \;-\; \frac{J_{0}}{\Omega_{0}}\;{\sf a}_{2}:\nabla{\bf u}. \]
We now use Eq.~\eqref{eq:S3_def} to obtain
\begin{eqnarray*}
S_{3} &=&  \frac{qB_{0}}{\Omega_{0}} \left( \Phi^{\prime}\,\Psi_{1}^{\prime} \;-\frac{}{} \Phi_{1}^{\prime}\right) \vb{\rho}\bdot\pd{\bf x}{\theta} \;-\; \frac{J_{0}}{\Omega_{0}}\;{\sf a}_{2}:\nabla{\bf u} \\
 &&+\; \frac{2}{3}\;\mu_{0}\,(B_{0}/\Omega_{0})\,\vb{\rho}\bdot\vb{\cal R},
 \end{eqnarray*}
which can be inserted into Eq.~\eqref{eq:G1zeta_3} to obtain
 \begin{eqnarray}
 G_{1}^{\zeta} &=& -\,\vb{\rho}\bdot\vb{\cal R} \;+\; \frac{{\sf a}_{2}}{\Omega_{0}}:\nabla{\bf u} \nonumber \\
  &&-\; q \left( \Phi^{\prime}\,\Psi_{1}^{\prime} \;-\frac{}{} \Phi_{1}^{\prime}\right) \pd{\vb{\rho}}{\mu_{0}}\bdot\pd{\bf x}{\theta}.
  \label{eq:G1zeta_gc}
 \end{eqnarray}
We note that the first term on the right side of Eq.~\eqref{eq:G1zeta_gc} is required to preserve gyrogauge invariance.
 
 \subsection{\label{sec:gc_Lag}Guiding-center Lagrangian in a drifting frame}
 
 By combining all relevant orders, and restoring the physical mass $\epsilon m \rightarrow m$, we construct the guiding-center Lagrangian in the drifting frame
 \begin{equation}
 L_{E{\rm gc}} \equiv \frac{q}{c}\,\Psi^{*}\dot{\Theta} + J\left(\dot{\zeta} - \dot{\Theta}\right) - \left( q\,\Phi^{*} + \mu\,B^{*}\right),
 \label{eq:LEgc_final}
 \end{equation}
 where 
 \begin{eqnarray}
 \Psi^{*}(\Psi) &\equiv& \Psi + \Psi_{1} \;=\; \Psi \left( 1 \;+\frac{}{} 2\,\epsilon\,\nu(\Psi)\right), \label{eq:Psi_star} \\
 \Phi^{*}(\Psi) &\equiv& \Phi + \Phi_{1} \;=\; \Phi(\Psi) \;+\; \epsilon\,\Psi\,\Phi^{\prime}(\Psi)\,\nu(\Psi), \label{eq:Phi_star} \\
 B^{*}(\Psi) &\equiv& B_{0} \left( 1 + \frac{1}{2}\,\Psi_{1}^{\prime}\right), \label{eq:B_star}
 \end{eqnarray}
 and $\epsilon = c\,\Phi^{\prime}(\Psi_{0})/\Omega_{0}$ returns to its physical interpretation, and $\nu(\Psi) \equiv \Phi^{\prime}(\Psi)/\Phi^{\prime}(\Psi_{0})$. The Euler-Lagrange guiding-center equations of motion for the guiding-center angles $\Theta$ and 
 $\zeta$ are
\begin{eqnarray}
\dot{\Theta} &=& \frac{c}{q} \left( q\,\frac{d\Phi^{*}}{d\Psi^{*}} + \mu\;\frac{dB^{*}}{d\Psi^{*}}\right) \equiv \Omega(\Psi,\mu), \label{eq:thetadot_gc} \\
\dot{\zeta} &=& \Omega_{0}\,B^{*}(\Psi)/B_{0} \;+\; \dot{\Theta}, \label{eq:eq:zetadot_gc}
\end{eqnarray}
where we note that the guiding-center azimuthal angular velocity \eqref{eq:thetadot_gc} depends on the guiding-center magnetic moment $\mu$ for nonlinear radial electric fields since $\Psi_{1}^{\prime\prime} \neq 0$. Since the guiding-center azimuthal angle $\Theta$ is ignorable, the guiding-center azimuthal canonical angular momentum 
 \begin{equation}
 P_{{\rm gc}\Theta} \;\equiv\; \pd{ L_{E{\rm gc}}}{\dot{\Theta}} \;=\; \frac{q}{c} \left( \Psi \;+\frac{}{} \Psi_{1} \right) \;-\; J
 \label{eq:Pgc_theta}
 \end{equation}
 is conserved, which follows from the conservation of $\Psi$ and $J$. It is also immediately clear that the guiding-center energy ${\cal E}_{\rm gc} \equiv q\Phi^{*}(\Psi) + \mu\,B^{*}(\Psi)$ is also a constant of motion. 
 
We note that the term $\frac{1}{2}\,\mu B_{0}\,\Psi_{1}^{\prime}$ in Eqs.~\eqref{eq:LEgc_final} and \eqref{eq:B_star} can be interpreted as a finite-Larmor-correction to the electrostatic potential energy
\[
q\,\langle\Phi({\bf X} + \vb{\rho})\rangle - q\,\Phi({\bf X}) = \frac{q}{2}\,\langle\vb{\rho}\vb{\rho}\rangle:\nabla\nabla\Phi = \frac{1}{2}\,\mu B_{0}\,\Psi_{1}^{\prime}.
\]
Hence, the guiding-center Hamiltonian can be expressed as
\begin{equation}
H_{\rm gc} \;=\; q\,\left\langle\Phi(\Psi - G_{1}^{\psi})\right\rangle \;+\; q\,\Phi_{1}(\Psi) \;+\; \mu\,B_{0},
\label{eq:Hgc_flr}
\end{equation}
where $q\,\Phi_{1} \equiv m\,|{\bf u}|^{2}/2$.

Lastly, we note that the guiding-center position can be expressed in Cartesian coordinates as $(X,Y)$, where
\begin{eqnarray}
X(t) &=& \sqrt{2\,\Psi/B_{0}}\;\cos\left[ \Omega(\Psi,\mu)\frac{}{} t\right], \label{eq:Xgc_dot} \\
Y(t) &=& \sqrt{2\,\Psi/B_{0}}\;\sin\left[ \Omega(\Psi,\mu)\frac{}{} t\right], \label{eq:Ygc_dot}
 \end{eqnarray}
 which can then be compared with the Cartesian coordinates $(x,y)$ of the particle position given by Eqs.~\eqref{eq:x_action}-\eqref{eq:y_action}. Hence, because of the conservation law of $\Psi$, the guiding-center moves on a circle with constant radius
 $\sqrt{2\,\Psi/B_{0}}$, at a constant angular velocity $\Omega(\Psi,\mu)$.
 
 \subsection{Guiding-center conservation laws}
 
 We have just discovered that the guiding-center motion conserves the guiding-center magnetic flux $\Psi$ and the guiding-center magnetic moment $\mu$. First, the guiding-center magnetic flux $\Psi$ can be constructed from the particle dynamics directly from the expansion
 \begin{equation}
\Psi \;=\; \psi \;+\; G_{1}^{\psi} \;+\; G_{2}^{\psi} \;+\; \frac{1}{2}\;G_{1}^{\beta}\pd{G_{1}^{\psi}}{z^{\beta}} + \cdots
\label{eq:Psi_gc}
\end{equation}
In Eq.~\eqref{eq:Psi_gc}, we find
\begin{eqnarray}
G_{1}^{\beta}\pd{G_{1}^{\psi}}{z^{\beta}} &=& \frac{B_{0}}{\Omega_{0}} \left[G_{1}^{\bf x}\bdot\nabla{\bf w}\bdot\pd{\bf x}{\theta} \;+\; G_{1}^{\bf x}\bdot\nabla\left(\pd{\bf x}{\theta}\right)\bdot{\bf w} \right] \nonumber \\
 &&+\; \frac{B_{0}}{\Omega_{0}} \left( G_{1}^{\mu}\,\pd{\bf w}{\mu_{0}} \;+\; G_{1}^{\zeta}\,\pd{\bf w}{\zeta_{0}}\right)\bdot\pd{\bf x}{\theta}.
 \end{eqnarray}
 Here, using $G_{1}^{\bf x} = -\,\vb{\rho}$, we find
 \[ G_{1}^{\bf x}\bdot\nabla{\bf w}\bdot\pd{\bf x}{\theta} \;=\; (\vb{\rho}\bdot\vb{\cal R})\;\pd{\bf w}{\zeta_{0}}\bdot\pd{\bf x}{\theta}, \]
 while
 \[ G_{1}^{\bf x}\bdot\nabla\left(\pd{\bf x}{\theta}\right)\bdot{\bf w} \;=\; \wh{\sf z}\bdot\left({\bf w}\btimes\vb{\rho}\right) \;=\; \frac{|{\bf w}|^{2}}{\Omega_{0}} \;=\; \frac{2J_{0}}{m}, \]
where $J_{0} = \mu_{0} B_{0}/\Omega_{0}$ is the lowest-order gyroaction, so that
 \[ \frac{1}{2}\,G_{1}^{\beta}\pd{G_{1}^{\psi}}{z^{\beta}} \;=\; \frac{c}{q}\,J_{0} \;+\; \frac{B_{0}}{2\Omega_{0}} \left( G_{1}^{\mu}\,\pd{\bf w}{\mu_{0}} \;+\; g_{1}^{\zeta}\,\pd{\bf w}{\zeta_{0}}\right)\bdot\pd{\bf x}{\theta}, \]
where $g_{1}^{\zeta} = G_{1}^{\zeta} + \vb{\rho}\bdot\vb{\cal R}$. Since $G_{2}^{\psi}$, given by Eq.~\eqref{eq:G2_psi}, is
\[ G_{2}^{\psi} = -\,\Psi_{1}^{\prime}\;G_{1}^{\psi} - \frac{B_{0}}{2\Omega_{0}}\left( G_{1}^{\mu}\,\pd{\bf w}{\mu_{0}} + g_{1}^{\zeta}\pd{\bf w}{\zeta_{0}}\right)\vb{\cdot}\pd{\bf x}{\theta}, \]
then 
\[ G_{2}^{\psi} \;+\; \frac{1}{2}\;G_{1}^{\beta}\pd{G_{1}^{\psi}}{z^{\beta}} \;=\; -\,\Psi_{1}^{\prime}\;G_{1}^{\psi} \;+\; \frac{c}{q}\,J_{0}. \]
Hence, the guiding-center magnetic flux $\Psi$ is defined as
\begin{equation}
\Psi \;=\; \psi \;+\; \left( 1 \;-\; \Psi_{1}^{\prime}\right) G_{1}^{\psi} \;+\; \frac{c}{q}\,J_{0} \;+\; \cdots,
\label{eq:Psi_gc_final}
\end{equation}
where $\Psi_{1}^{\prime} = 2\,c(\Phi^{\prime} + \psi\;\Phi^{\prime\prime})/\Omega_{0}$, and $G_{1}^{\psi} = 2\psi\,(\dot{\theta} - c\Phi^{\prime}/\Omega_{0})$. We also note that the gyroangle-averaged magnetic flux $\langle\psi\rangle = \Psi - (c/q)\,J \neq \Psi$ is not equal to the guiding-center magnetic flux.

Next, the guiding-center magnetic moment $\mu_{\rm gc}$ can be constructed from the particle dynamics directly from the expansion
\begin{equation}
\mu \;=\; \mu_{0} \;+\; G_{1}^{\mu} + \cdots,
\label{eq:mu_gc}
\end{equation}
where the lowest-order magnetic moment $\mu_{0} = m|{\bf w}|^{2}/2B_{0}$ is
\begin{equation}
\mu_{0} \;=\; \frac{q\;\psi }{c\Omega_{0}B_{0}}\left[ \left(\frac{\dot{\psi}}{2\psi}\right)^{2} + \left(\dot{\theta} - c\Phi^{\prime}\right)^{2} \right],
\label{eq:mu0}
\end{equation}
and 
\begin{eqnarray}
G_{1}^{\mu} &=& -\;\frac{\mu_{0}}{\Omega_{0}}\left({\sf a}_{1}:\nabla{\bf u} \;+\frac{}{} \wh{\sf z}\bdot\nabla\btimes{\bf u} \right) \nonumber \\
 &&+ \frac{q}{B_{0}} \left( \Phi^{\prime}\,\Psi_{1}^{\prime} \;-\frac{}{} \Phi_{1}^{\prime}\right) G_{1}^{\psi},
\end{eqnarray}
with
\begin{eqnarray*} 
\Phi^{\prime}\,\Psi_{1}^{\prime} - \Phi_{1}^{\prime} &=& \frac{c\Phi^{\prime}}{\Omega_{0}} \left[(2\,\Phi^{\prime} + 2\psi\;\Phi^{\prime\prime}) \;-\frac{}{} (\Phi^{\prime} + 2\psi\;\Phi^{\prime\prime})\right] \nonumber \\
 &=& \frac{\Omega_{0}}{c} \left( \frac{c\Phi^{\prime}}{\Omega_{0}}\right)^{2}.
 \end{eqnarray*}
Here, we use Eq.~\eqref{eq:a1_eq} to write
\[  -\,\frac{\mu_{0}}{\Omega_{0}}\;{\sf a}_{1}:\nabla{\bf u} \;=\; \frac{m}{2B_{0}}\;({\bf w}\vb{\rho}):\nabla{\bf u} \;+\; \frac{\mu_{0}\wh{\sf z}}{2\Omega_{0}}\bdot\nabla\btimes{\bf u}, \]
where
\begin{eqnarray}
 \frac{m}{2B_{0}}\;({\bf w}\vb{\rho}):\nabla{\bf u} &=& -\,\frac{q\psi}{B_{0}}\;\left(\Phi^{\prime} + 2\psi\,\Phi^{\prime\prime}\right)\left(\dot{\theta} - \frac{c\Phi^{\prime}}{\Omega_{0}}\right)^{2} \nonumber \\
  &&-\; \frac{q\psi}{B_{0}}\;\Phi^{\prime} \left(\frac{\dot{\psi}}{2\psi}\right)^{2},
 \end{eqnarray}
 and
 \begin{equation} 
 \frac{\mu_{0}\wh{\sf z}}{2\Omega_{0}}\bdot\nabla\btimes{\bf u} = \frac{1}{2}\,\mu_{0}\,\Psi_{1}^{\prime} \;=\; \mu_{0} \left( \Phi^{\prime} + \psi\;\Phi^{\prime\prime}\right).
  \end{equation} 
  We thus easily conclude that, from Eq.~\eqref{eq:mu_gc}, we find the simple relation $\langle\mu_{0}\rangle = \mu - \langle G_{1}^{\mu}\rangle \equiv \mu\,d\Psi^{*}/d\Psi$. The conservation laws of the guiding-center azimuthal canonical angular momentum \eqref{eq:Psi_gc_final} and the guiding-center magnetic moment \eqref{eq:mu_gc} will be explored in Sec.~\ref{sec:linear_gc} for the case of a linear radial electric field.
  
  Lastly, we establish the validity of the guiding-center representation by verifying that the guiding-center pull-back ${\sf T}_{\rm gc}P_{{\rm gc}\Theta}$ of the guiding-center azimuthal canonical angular momentum \eqref{eq:Pgc_theta} is equal to the particle azimuthal canonical angular momentum \eqref{eq:p_theta}:  ${\sf T}_{\rm gc}P_{{\rm gc}\Theta} = p_{\theta}$. Here, the expansion of the guiding-center pull-back ${\sf T}_{\rm gc}P_{{\rm gc}\Theta}$
  \begin{eqnarray}
 {\sf T}_{\rm gc}P_{{\rm gc}\Theta} &=& \frac{q}{c}\,\psi \;+\; \epsilon\,\frac{q}{c} \left( \Psi_{1} + G_{1}^{\psi}\right) \;-\; \epsilon^{2}\;J \nonumber \\
  &&+\; \epsilon^{2}\frac{q}{c}\left( G_{2}^{\psi} + \frac{1}{2}\,{\sf G}_{1}\cdot\exd G_{1}^{\psi} + G_{1}^{\psi}\,\Psi_{1}^{\prime}\right) \nonumber \\
   &=& \frac{q}{c}\,\psi \left(1 \;+\frac{}{} 2\,\dot{\theta}/\Omega_{0}\right) \;=\; p_{\theta}
 \end{eqnarray}
yields the particle azimuthal canonical angular momentum $p_{\theta}$ up to second order in $\epsilon$. Hence, the guiding-center transformation \eqref{eq:Zz_gc} generated by the components $(G_{1}^{\alpha},G_{2}^{\alpha},...)$ is faithful to the exact conservation laws of the particle dynamics.

\subsection{Guiding-center polarization and magnetization}

Polarization and magnetization are pillars of the reduced Vlasov-Maxwell dynamical description of self-consistent magnetized plasmas \cite{Brizard:2008,Brizard:2013,Tronko_Brizard:2015}. We now calculate the guiding-center polarization and magnetization in the lab frame, which are each defined as the sum of a contribution associated with the transformation to the drifting frame and a contribution in the drifting frame directly calculated from the guiding-center transformation. 

We begin with the guiding-center polarization, which is expressed in terms of the electric-dipole definition
\begin{equation}
\vb{\pi}_{\rm gc} \;\equiv\; q\,\left(\vb{\rho}_{E} \;+\frac{}{} \langle\vb{\rho}_{\rm gc}\rangle\right), 
\label{eq:pi_gc_def}
\end{equation}
where the lowest-order electric displacement 
\begin{equation}
\vb{\rho}_{E} \;\equiv\; \frac{\wh{\sf z}}{\Omega_{0}}\btimes{\bf u} \;=\; -\;\frac{c\,\nabla\Phi}{B_{0}\Omega_{0}} \;=\; -\;\frac{c\,\Phi^{\prime}(\psi)}{B_{0}\Omega_{0}}\;\nabla\psi
\label{eq:rho_E}
\end{equation}
involves the radial electric field, as expected. The contribution associated with the guiding-center transformation is constructed from the guiding-center displacement 
$\vb{\rho}_{\rm gc} \equiv {\sf T}_{\rm gc}^{-1}{\bf x} - {\bf X}$, which is expressed as
\begin{eqnarray*}
\vb{\rho}_{\rm gc} &=& -\,\epsilon\,G_{1}^{\bf x} \;-\; \epsilon^{2} \left( G_{2}^{\bf x} \;-\; \frac{1}{2}\,{\sf G}_{1}\cdot\exd G_{1}^{\bf x} \right) \;+\; \cdots \nonumber \\
 &=& \epsilon\left(1 - \epsilon\,\Psi_{1}^{\prime}\right)\,\vb{\rho} - \frac{\epsilon^{2}\wh{\sf z}}{\Omega_{0}}\btimes\left( G_{1}^{\mu}\pd{\bf w}{\mu} + g_{1}^{\zeta}\pd{\bf w}{\zeta}\right),
\end{eqnarray*}
where we have restored the mass renormalization $m \rightarrow \epsilon\,m$. Given the fact that the lowest-order gyroradius $\vb{\rho}$ is gyroangle-dependent, the gyroangle-averaged guiding-center displacement yields the expression
\begin{eqnarray}
\langle\vb{\rho}_{\rm gc}\rangle &=& -\;\frac{\epsilon^{2}\wh{\sf z}}{\Omega_{0}}\btimes\left\langle G_{1}^{\mu}\pd{\bf w}{\mu} + g_{1}^{\zeta}\pd{\bf w}{\zeta}\right\rangle \nonumber \\
 &=& \frac{\epsilon^{2}c}{B_{0}\Omega_{0}} \left( \Phi^{\prime}\,\Psi_{1}^{\prime} \;-\frac{}{} \Phi_{1}^{\prime}\right) \nabla\psi,
 \label{eq:rho_gc_av}
\end{eqnarray}
where we used Eqs.~\eqref{eq:G1mu_3}-\eqref{eq:G1zeta_3}. By adding the two contributions \eqref{eq:rho_E} and \eqref{eq:rho_gc_av} in Eq.~\eqref{eq:pi_gc_def}, we find the net guiding-center electric-dipole moment
\begin{equation}
\vb{\pi}_{\rm gc} \;=\; -\;\frac{cq}{B_{0}\Omega_{0}} \left[ \Phi^{\prime} \;+\frac{}{} \epsilon \left(\Phi_{1}^{\prime} - \Phi^{\prime}\Psi_{1}^{\prime}\right) \right] \nabla\psi,
\label{eq:pi_gc_final}
\end{equation}
which contains first-order guiding-center corrections to the lowest-order electric displacement. We now show that, using Eq.~\eqref{eq:Xgc_dot}, Eq.~\eqref{eq:pi_gc_final} can be expressed as
\begin{eqnarray}
\vb{\pi}_{\rm gc} &\equiv& \frac{q\wh{\sf z}}{\Omega_{0}}\btimes\dot{\bf X} \;=\; -\;\frac{cq\,\nabla\Phi^{*}}{B_{\|}^{*}\Omega_{0}} \;=\;  -\;\frac{cq}{B_{0}\Omega_{0}} \;\frac{d\Phi^{*}}{d\Psi^{*}}\,\nabla\psi \nonumber \\
 &=& -\;\frac{cq}{B_{0}\Omega_{0}} \left(\frac{\Phi^{\prime} + \epsilon\,\Phi_{1}^{\prime}}{1 \;+\; \epsilon\,\Psi_{1}^{\prime}} \right) \nabla\psi,
 \label{eq:Xdot_pol}
\end{eqnarray}
which yields Eq.~\eqref{eq:pi_gc_final} if we explicitly expand Eq.~\eqref{eq:Xdot_pol} in powers of $\epsilon$ and keep only terms up to second order. 

We note that the drifting-frame guiding-center polarization contribution can also be calculated from the 
guiding-center Lagrangian \eqref{eq:LEgc_final}, which can be rewritten as
\begin{eqnarray}
L_{E{\rm gc}} &=& \left(\frac{q}{c}\,\Psi\,\nabla\Theta \;+\; m\,{\bf u}\right)\bdot\dot{\bf X} + J\,\left(\dot{\zeta} - \vb{\cal R}\bdot\dot{\bf X}\right) \\
 &&-\; \left( q\,\Phi \;+\; \frac{m}{2}\,|{\bf u}|^{2} \;+\; \mu\,B_{0} \;-\; \frac{q}{2}\,\langle\vb{\rho}\vb{\rho}\rangle:\nabla{\bf E}\right). \nonumber 
 \end{eqnarray}
 Hence, we find
 \begin{equation}
 \pd{L_{E{\rm gc}}}{\bf E}  \;=\; \frac{q\wh{\sf z}}{\Omega_{0}}\btimes\dot{\bf X} \;-\; \frac{qc\,{\bf E}}{B_{0}\Omega_{0}} \;=\; \frac{q\wh{\sf z}}{\Omega_{0}}\btimes\dot{\bf X} - q\,\vb{\rho}_{E},
 \end{equation}
while the quadrupole contribution
 \[ \nabla\bdot \left(\pd{L_{E{\rm gc}}}{\nabla{\bf E}}\right) \;=\; \frac{q}{2}\;\nabla\bdot\langle\vb{\rho}\vb{\rho}\rangle \;=\; 0 \]
 vanishes in a uniform magnetic field.

Next, we calculate the guiding-center intrinsic magnetic dipole
\begin{eqnarray}
\vb{\mu}_{\rm gc} &\equiv& \frac{q\Omega_{0}}{2c} \left\langle \vb{\rho}_{\rm gc}\btimes\pd{\vb{\rho}_{\rm gc}}{\zeta}\right\rangle \\
 &=& \frac{q\Omega_{0}}{2c} \left( 1 \;-\frac{}{} 2\,\epsilon\,\Psi_{1}^{\prime}\right) \left\langle \vb{\rho}\btimes\pd{\vb{\rho}}{\zeta}\right\rangle \nonumber \\
  &&-\; \frac{\epsilon q}{c} \left\langle \left[\wh{\sf z}\btimes\left( G_{1}^{\mu}\pd{\bf w}{\mu} + g_{1}^{\zeta}\pd{\bf w}{\zeta}\right)\right]\btimes\pd{\vb{\rho}}{\zeta}\right\rangle. \nonumber 
\end{eqnarray}
The lowest-order contribution makes use of the definition $\langle \vb{\rho}\btimes\partial\vb{\rho}/\partial\zeta\rangle = - (2\mu_{0}B_{0}/m\Omega_{0}^{2})\,\wh{\sf z}$, so that we find
\[ \frac{q\Omega_{0}}{2c} \left( 1 \;-\frac{}{} 2\,\epsilon\,\Psi_{1}^{\prime}\right) \left\langle \vb{\rho}\btimes\pd{\vb{\rho}}{\zeta}\right\rangle \;=\; -\;\mu_{0}\left( 1 \;-\frac{}{} 2\,\epsilon\,\Psi_{1}^{\prime}\right)\,\wh{\sf z}, \]
while the first-order contribution is
\[ \frac{\epsilon q}{c} \left\langle \left[\wh{\sf z}\btimes\left( G_{1}^{\mu}\pd{\bf w}{\mu} + g_{1}^{\zeta}\pd{\bf w}{\zeta}\right)\right]\btimes\pd{\vb{\rho}}{\zeta}\right\rangle \;=\; -\;\epsilon\,\mu_{0}\Psi_{1}^{\prime}\,\wh{\sf z}. \]
If we combine these results, we obtain the simple formula
\begin{equation}
\vb{\mu}_{\rm gc} \;\equiv\; -\;\mu_{0}\;(B_{0}/B_{\|}^{*})\,\wh{\sf z} \;=\; -\;\mu_{0}\left( 1 \;-\frac{}{} \epsilon\,\Psi_{1}^{\prime}\right)\,\wh{\sf z},
\end{equation}
after an expansion in powers of $\epsilon$ is carried out.
 
 \section{\label{sec:linear_gc}Guiding-center Dynamics for a Linear Radial Electric Field}
 
 In this Section, we return to the case of a linear radial electric field, where $\Phi^{\prime} = \epsilon\,\Omega_{0}/c$ and $\Phi^{\prime\prime} = 0$, so that $\epsilon\Psi_{1}^{\prime} = 2\epsilon$ and $\epsilon\Phi_{1}^{\prime} = \epsilon^{2}\Omega_{0}/c$. In this case, the guiding-center azimuthal angular velocity \eqref{eq:thetadot_gc} is $\Omega(\Psi) = \Omega_{0}\,\ov{\omega}(\epsilon)$, where $\ov{\omega}(\epsilon) = \epsilon\,(1 + \epsilon)/(1 + 2 \epsilon)$, i.e., in the limit $\epsilon \ll 1$, the guiding-center azimuthal angular velocity is proportional to $\epsilon = c\,\Phi^{\prime}(\ov{\psi}_{0})/\Omega_{0}$. As was noted below Eq.~\eqref{eq:thetadot_gc}, the guiding-center azimuthal angular velocity $\Omega(\Psi)$ is independent of the guiding-center magnetic moment $\mu$ for a linear radial electric field since $\Psi_{1}^{\prime}$ is a constant.
  
 Here, we will use the dimensionless ordering parameter $\epsilon = 1/30$, instead of the standard value $1/1000$ commonly assumed in guiding-center theory, in order to show how far the perturbation analysis can be pushed to nonstandard orderings, e.g., according to Joseph's ordering \cite{Joseph:2021}, we find $\Omega_{E}/\Omega_{0} = \sqrt{2\epsilon} \simeq 25\%$.
 
 \begin{figure}
\epsfysize=1.8in
\epsfbox{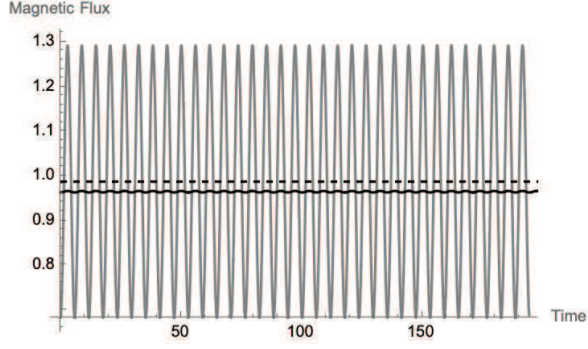}
\caption{Plots of the normalized magnetic flux $\chi(\tau) \equiv \psi(\tau)/\ov{\psi}_{0}$ (gray) and the normalized guiding-center magnetic flux $\chi_{\rm gc}(\tau) \equiv \Psi(\tau)/\ov{\psi}_{0}$ (black) for the case of a linear radial electric field with $\epsilon = 1/30$ and $\phi = \pi/10$. The horizontal dashed  line corresponds to the averaged magnetic flux $\langle\chi\rangle = (a^{2} + b^{2})/2 = \chi_{0}(\epsilon)\,\sec\phi$.}
\label{fig:chi_gc}
\end{figure}

  \begin{figure}
\epsfysize=2in
\epsfbox{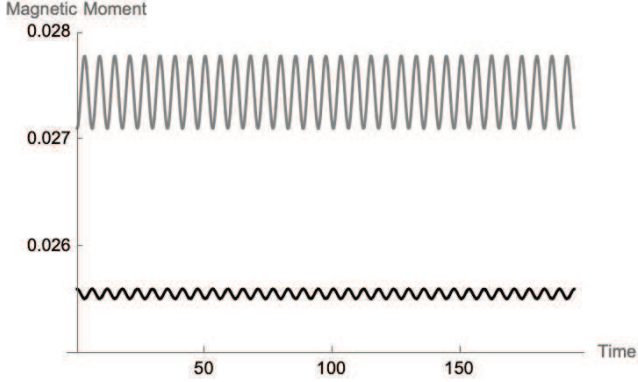}
\caption{Plots of the normalized magnetic moment $\ov{\mu}_{0}(\tau)$ (gray) and the normalized guiding-center magnetic moment $\ov{\mu}(\tau)$ (black) for the case of a linear radial electric field with $\epsilon = 1/30$ and $\phi = \pi/10$. }
\label{fig:mu_gc}
\end{figure}

\subsection{Guiding-center conservation laws}

In the case of a linear radial electric field, the guiding-center magnetic flux \eqref{eq:Psi_gc_final} becomes
 \begin{eqnarray}
 \Psi(\tau) &=& (1 - 2\epsilon)\,\ov{\psi}_{0} \;+\; \frac{c}{q\Omega_{0}}\left( {\cal E} \;-\frac{}{} q\,\Phi_{0}\right) \nonumber \\
  &&+\; 5\,\epsilon^{2}\,\psi(\tau), 
  \label{eq:Psi_linear}
 \end{eqnarray}
 where  $\Phi_{0} \equiv \Phi(\ov{\psi}_{0})$ and the time dependence (with $\tau = \sqrt{1 + 4 \epsilon}\,\Omega_{0}\,t$) has been pushed from zeroth order to second order in $\epsilon$ as a result of the guiding-center transformation \eqref{eq:Psi_gc_final}. Next, the guiding-center magnetic moment \eqref{eq:mu_gc} becomes
\begin{eqnarray}
\mu(\tau) &=& (1 - 2\,\epsilon)\;\frac{\cal E}{B_{0}} \;-\; (1 - 3\,\epsilon)\;\frac{q\Phi_{0}}{B_{0}} \nonumber \\
 &&-\; 4\,\epsilon^{3}\left(\frac{q\Omega_{0}}{cB_{0}}\right)\psi(\tau),
 \label{eq:mu_linear}
\end{eqnarray}
where $\mu_{0}$ is given in Eq.~\eqref{eq:mu0}: 
\begin{equation} 
\mu_{0} \;=\; ({\cal E} - q\,\Phi_{0})/B_{0} \;+\; \epsilon^{2}\left(\frac{q\Omega_{0}}{cB_{0}}\right)\psi(\tau).
\label{eq:mu0_linear}
\end{equation}
Hence, the time dependence has been pushed from second order to third order in $\epsilon$ as a result of the guiding-center transformation \eqref{eq:mu_gc}.

In Fig.~\ref{fig:chi_gc}, we see the normalized lowest-order magnetic flux $\chi(\tau) \equiv \psi(\tau)/\ov{\psi}_{0}$ (gray) and the normalized guiding-center magnetic flux $\chi_{\rm gc}$ (black) for the case of a linear radial electric field with $\epsilon = 1/30$ and $\phi = \pi/10$. We clearly see that the large-amplitude oscillation in $\psi(\tau)$ has been greatly reduced in Eq.~\eqref{eq:Psi_linear} by a factor of $\epsilon^{2}$. We also see that the normalized guiding-center magnetic flux $\chi_{\rm gc}$ is NOT equal to the averaged normalized magnetic flux $\langle\chi\rangle = (a^{2} + b^{2})/2 = \chi_{0}\sec\phi$, shown as a dashed horizontal line in Fig.~\ref{fig:chi_gc}. 

In Fig.~\ref{fig:mu_gc}, we see the normalized lowest-order magnetic moment $\ov{\mu}_{0}$ (gray) and the normalized guiding-center magnetic moment $\ov{\mu}$ (black) for the case of a linear radial electric field with $\epsilon = 1/30$ and $\phi = \pi/10$, each normalized by $(q\Omega_{0}/cB_{0})\ov{\psi}_{0}$. We clearly see that, while the lowest-order magnetic moment \eqref{eq:mu0_linear} shows some oscillations with small amplitudes (at order $\epsilon^{2}$), the guiding-center magnetic moment \eqref{eq:mu_linear} is fairly flat, with minimal-amplitude oscillations (at order $\epsilon^{3}$). 

\subsection{Guiding-center dynamics}

 \begin{figure}
\epsfysize=1.8in
\epsfbox{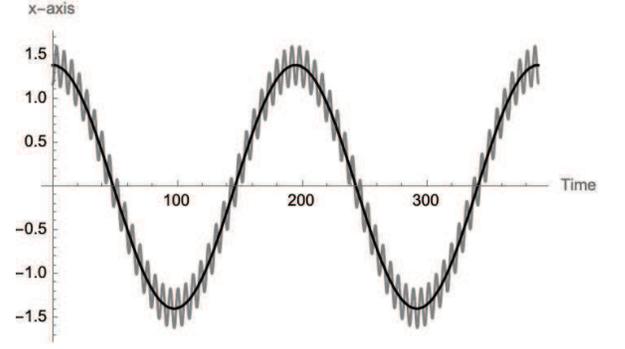}
\caption{Plots of $x(t')$ and $X(t')$ shown as gray and black curves, respectively,  in the range $0 \leq t' \leq 4\pi/\ov{\omega}(\epsilon)$ for the case of a linear radial electric field with $\epsilon = 1/30$ and $\phi = \pi/10$.}
\label{fig:x_gc}
\end{figure}

 \begin{figure}
\epsfysize=2.8in
\epsfbox{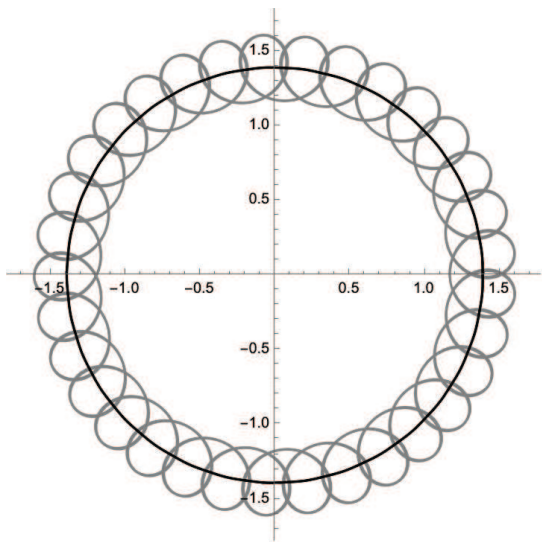}
\caption{Parametric plots of $(x,y)$ and $(X,Y)$ shown as gray and black curves, respectively, in the range $0 \leq t' \leq 2\pi/\ov{\omega}(\epsilon)$ for the case of a linear radial electric field with $\epsilon = 1/30$ and $\phi = \pi/10$.}
\label{fig:xy_gc}
\end{figure}

Lastly, the plots of $x(t')$ and $X(t')$, as well as the parametric plots of $(x,y)$ and $(X,Y)$, are shown in Figs.~\ref{fig:x_gc} and \ref{fig:xy_gc}, respectively, for the case of a linear radial electric field with $\epsilon = 1/30$ and $\phi = \pi/10$. We clearly see how well the guiding-center position \eqref{eq:Xgc_dot}-\eqref{eq:Ygc_dot}:
\begin{eqnarray}
X(t') &=& \sqrt{2\,\Psi(\tau)/B_{0}}\;\cos\left(t^{\prime}\frac{}{}\ov{\omega}(\epsilon)\right), \\
Y(t') &=& \sqrt{2\,\Psi(\tau)/B_{0}}\;\sin\left(t^{\prime}\frac{}{}\ov{\omega}(\epsilon)\right) 
\end{eqnarray}
follows the particle position \eqref{eq:x_final}-\eqref{eq:y_final}. Hence, the guiding-center transformation introduced in Sec.~\ref{sec:nonlinear_gc} has achieved its purpose in building guiding-center invariants $\Psi$ and $\mu$ to higher order in perturbation analysis from the lowest-order coordinates $\psi$ and $\mu_{0}$. In addition, the guiding-center dynamics follows the particle dynamics 

\section{Summary}

The presence of a nonuniform electric field adds a significant element of complexity in the guiding-center analysis of particle motion in crossed electric and magnetic fields, which are quite common in laboratory and space magnetized plasmas. In the present work, we greatly simplified the guiding-center analysis presented in Ref.~\cite{Brizard:1995} by considering a nonuniform radial electric field in the presence of a uniform magnetic field. 

The case of a nonlinear radial electric field is a topic of great interest in the investigation of turbulence and transport in rotating magnetized plasmas \cite{Hahm:1996,Hahm:2002,Peeters:2009,Sugama:2011,Belli_Candy:2018} and was recently explored by Wang {\it et al.} \cite{Wang_etal:2021} in performing gyrokinetic studies of ion-temperature-gradient (ITG) turbulence and transport in the scrape-off layer (SOL) region of a field-reversed magnetized plasma. 

The results of our guiding-center analysis for the case of a linear radial electric field confirm the faithfulness of the guiding-center representation. For a nonlinear radial electric field with quadratic nonlinearity
\[ \Phi(\psi) \;=\; \Phi_{0} \;+\; \Phi^{\prime}_{0}\,(\psi - \ov{\psi}_{0}) \;+\; \frac{1}{2}\,\Phi^{\prime\prime}_{0}\,(\psi - \ov{\psi}_{0})^{2}, \]
the radial integral solution \eqref{eq:t_chi} involves Weierstrass elliptic functions (for example, see Ref.~\cite{Brizard:2015}), and the energy dependence of the guiding-center azimuthal angular velocity \eqref{eq:thetadot_gc} becomes important.  Additional comments concerning guiding-center orbits in a nonlinear radial electric field in a uniform magnetic field can be found in the recent work by Joseph \cite{Joseph:2021}.  Future work will consider other orbital effects of nonlinear radial electric field such as the orbit squeezing effect \cite{Hazeltine_1989,Shaing_Hazeltine_1992,Shaing_1997}, which may be explored in the limit of a uniform magnetic field, as well as applications of the general guiding-center theory presented in Sec.~\ref{sec:nonlinear_gc} for the case of a nonlinear radial electric field in a nonuniform magnetic field.

\vspace*{0.2in}
\acknowledgments

The Author wishes to acknowledge useful discussions with K.~Kabin on the charged-particle dynamics in the presence of a linear radial electric field. The present work was supported by the National Science Foundation grant PHY-2206302.

\vspace*{0.1in}

\begin{center}
{\bf Data Availability Statement}
\end{center}

The Mathematica code used to generate the plots in the present manuscript is available upon request.


\begin{thebibliography}{99} 

\bibitem{Hahm:1996} T.S.~Hahm, Phys.~Plasmas {\bf 3}, 4658 (1996).

\bibitem{Hahm:2002} T.S.~Hahm, Plasma Phys.~Control.~Fusion {\bf 44}, A87 (2002).

\bibitem{Peeters:2009} A.G. Peeters, D. Strintzi, Y. Camenen, C. Angioni, F. J. Casson, W. A. Hornsby, and A. P. Snodin, Phys.~Plasmas {\bf 16}, 042310 (2009).

\bibitem{Burrell:2020} K.H.~Burrell, Phys.~Plasmas {\bf 27}, 060501 (2020).

\bibitem{Brizard:1995} A.J.~Brizard, Phys.~Plasmas {\bf 2}, 459 (1995).

\bibitem{Cary_Brizard:2009} J.R.~Cary and A.J.~Brizard, Rev.~Mod.~Phys.~{\bf 81}, 693 (2009).

\bibitem{Dimits:2010} A.M.~Dimits, Phys.~Plasmas {\bf 17}, 055901 (2010).

\bibitem{Sugama:2011} H.~Sugama, T. H. Watanabe, M. Nunami, and S. Nishimura, Plasma Phys.~Control.~Fusion {\bf 53}, 024004 (2011).

\bibitem{Belli_Candy:2018} E.~A.~Belli and J.~Candy, Phys.~Plasmas {\bf 25}, 032301 (2018).

\bibitem{Frei_etal:2020} B.J.~Frei, R.~Jorge, and P.~Ricci, J.~Plasma Phys.~{\bf 86}, 905860205 (2020).

\bibitem{Wang_etal:2021} W.H. Wang, J. Bao, X. S. Wei, Z. Lin, G. J. Choi, S. Dettrick, A. Kuley, C. Lau, P. F. Liu, and T. Tajima, Plasma Phys.~Control.~Fusion {\bf 63}, 065001 (2021).

\bibitem{White:2018} R.B.~White, A.~Hassam, and A.~Brizard, Phys.~Plasmas {\bf 25}, 012514 (2018).

\bibitem{Joseph:2021} I.~Joseph, Phys.~Plasmas {\bf 28}, 042102 (2021).

\bibitem{Brizard:2015}  A.J.~Brizard, {\it An Introduction to Lagrangian Mechanics}, 2nd ed.~(World Scientific, 2015). 

\bibitem{Kabin:2023} K.~Kabin, private communication (2023).

\bibitem{Littlejohn:1982} R.G.~Littlejohn, J.~Math.~Phys.~{\bf 23}, 742 (1982).

\bibitem{Brizard:2008} A.J.~Brizard, Comm.~Nonlin.~Sci.~Num.~Sim.~{\bf 13}, 24 (2008).

\bibitem{Brizard:2013} A.J.~Brizard, Phys.~Plasmas {\bf 20}, 092309 (2013).

\bibitem{Tronko_Brizard:2015} N.~Tronko and A.J.~Brizard, Phys.~Plasmas {\bf 22}, 112507 (2015).

\bibitem{Hazeltine_1989} R.D. Hazeltine, Phys. Fluids {\bf B1}, 2031 (1989).

\bibitem{Shaing_Hazeltine_1992} K.C. Shaing and R.D. Hazeltine, Phys. Fluids {\bf B4}, 2547 (1992).

\bibitem{Shaing_1997} K.C. Shaing and M.C. Zarnstorff, Phys. Plasmas {\bf 4}, 3928 (1997).


\end{thebibliography}
\end{document}